\input harvmac
\input epsf
\def\journal#1&#2(#3){\unskip, \sl #1\ \bf #2 \rm(19#3) }
\def\andjournal#1&#2(#3){\sl #1~\bf #2 \rm (19#3) }

\def\ie{{\it i.e.}}
\def\eg{{\it e.g.}}

\def\frac#1#2{{#1\over#2}}

\def\inbar{\,\vrule height1.5ex width.4pt depth0pt}
\def\IC{\relax\hbox{$\inbar\kern-.3em{\rm C}$}}
\def\IR{\relax{\rm I\kern-.18em R}}
\def\IP{\relax{\rm I\kern-.18em P}}
\def\IZ{\relax{\rm I\kern-.18em Z}}

%
%
\def\np#1#2#3{Nucl. Phys. {\bf B#1} (#2) #3}

\catcode`\@=11
\def\slash#1{\mathord{\mathpalette\c@ncel{#1}}}
\overfullrule=0pt

\def\underrel#1\over#2{\mathrel{\mathop{\kern\z@#1}\limits_{#2}}}

\catcode`\@=12


%

\def\exp{{\rm exp}}


\rightline{RI-02-00, EFI-2000-7, CERN-TH/2000-095}
\Title{
\rightline{hep-th/0005052}}
{\vbox{\centerline{D-Branes in the Background of NS Fivebranes}}}
\medskip
\centerline{\it Shmuel Elitzur${}^{1}$, Amit Giveon${}^{1,2}$,
David Kutasov${}^{3}$, Eliezer Rabinovici${}^{1}$ and Gor Sarkissian${}^{1}$}
\bigskip
\smallskip
\centerline{${}^1$Racah Institute of Physics, The Hebrew University}
\centerline{Jerusalem 91904, Israel}
\smallskip
\centerline{${}^2$Theory Division, CERN}
\centerline{CH-1211, Geneva 23, Switzerland}
\smallskip
\centerline{${}^3$Department of Physics, University of Chicago}
\centerline{5640 S. Ellis Av., Chicago, IL 60637, USA }
\bigskip\bigskip\bigskip
\noindent
We study the dynamics of $D$-branes in the near-horizon geometry of $NS$
fivebranes. This leads to a holographically dual description of the
physics of $D$-branes ending on and/or intersecting $NS5$-branes. We
use it to verify some properties of such $D$-branes which were deduced
indirectly in the past, and discuss some instabilities of non-supersymmetric
brane configurations. Our construction also describes vacua of Little String
Theory which are dual to open plus closed string theory in asymptotically
linear dilaton spacetimes.
\vfill
\Date{5/00}
\newsec{Introduction}

\lref\gkrev{A. Giveon and D. Kutasov, hep-th/9802067,
Rev. Mod. Phys. {\bf 71} (1999) 983.}%

In the last few years it was found that embedding various
supersymmetric gauge theories in string theory, as the low
energy worldvolume dynamics on branes, provides an efficient
tool for studying many aspects of the vacuum structure and
properties of BPS states in these theories. One class of
constructions (reviewed in \gkrev) involves systems of $D$-branes
ending on and/or intersecting $NS5$-branes. For applications to
gauge theory one is typically interested in taking the weak coupling
limit $g_s\to 0$ (as well as the low energy limit). In this limit
one might expect the system to be amenable to a perturbative worldsheet
treatment, but the presence of the $NS5$-branes and branes ending on
branes complicate the analysis.

In the absence of a derivation of the properties of $D$-branes
interacting with $NS5$-branes from first principles, some of
their low energy properties were postulated in the past based
on symmetry considerations and consistency conditions. One of
the main purposes of this paper is to derive some of these
properties by a direct worldsheet study of $D$-branes in the
vicinity of $NS5$-branes.

\lref\ahar{For a review see O. Aharony, hep-th/9911147,
Class. Quant. Grav. {\bf 17} (2000) 929.}%
\lref\abks{O. Aharony, M. Berkooz, D. Kutasov and N. Seiberg,
hep-th/9808149, JHEP {\bf 9810} (1998) 004.}%

In the analysis, we will use the improved understanding of the
dynamics of $NS5$-branes achieved in the last few years. It is
now believed that in the weak coupling limit (but not necessarily
at low energies) fivebranes decouple from gravity and other bulk
string modes and give rise to a rich non-gravitational theory,
Little String Theory (LST) \ahar. In \abks\ it was proposed
to study LST using holography. The theory on a stack of $NS5$-branes
was argued to be holographically dual to string theory in the
near-horizon geometry of the fivebranes. Many properties of LST
can be understood by performing computations in string theory in
this geometry. In particular, we will find below that
it is an efficient way to study properties of $D$-branes ending on
or intersecting $NS5$-branes.

{}From the point of view of LST, $D$-branes in the vicinity of
$NS5$-branes give rise to new particle and extended object
states in the theory, as well as new vacua, typically with
reduced supersymmetry (when the branes are space-filling).
Another motivation of this work is to understand the
non-perturbative spectrum and dynamics of extended objects in
LST, and more generally study the interplay between the
non-trivial dynamics on fivebranes and the worldvolume physics
of $D$-branes in their vicinity.

\lref\chs{C. Callan, J. Harvey and A. Strominger, hep-th/9112030.}%

The plan of the paper is the following. In section 2 we briefly
review some facts regarding the relevant brane configurations and,
in particular, describe the conjectured properties of these
configurations that we will try to verify. Section 3 is a review
of the near-horizon geometry of $NS5$-branes (the CHS geometry
\chs) and its holographic relation to LST. We also describe a
modification of this geometry corresponding to fivebranes
positioned at equal distances around a circle, which plays a role
in the analysis. In section 4 we review some facts about $D$-branes
in flat space and on a three-sphere (or $SU(2)$ WZW model). In section
5 we study $D$-branes in the CHS geometry as well as its regularized
version. We verify some of the properties described in section 2, and
describe additional features which follow from our analysis.
Some of the technical details are presented in the appendices.

\newsec{Some properties of brane configurations}

One class of brane constructions, that gave rise to many insights
into gauge dynamics, involves $D$-branes suspended between $NS5$-branes.
We start with a brief description of a particular example of such a
construction, which realizes four dimensional $N=1$ supersymmetric
gauge theory with gauge group $G=U(N_c)$ and $N_f$ chiral superfields
in the fundamental representation (more precisely, $N_F$ fundamentals
$Q^i$, $i=1,\cdots, N_F$, and $N_F$ anti-fundamentals $\tilde Q_i$).
This will help introduce the issues that will be discussed
later\foot{For a more detailed discussion, references to the original
literature and other constructions, see \gkrev.}.

We will consider brane configurations in type IIA string theory
consisting of $NS5$-branes, $D4$ and $D6$-branes, oriented as
follows\foot{In later sections we will discuss more general
configurations, obtained by rotating some of the branes.}:
\eqn\nsd{\eqalign{
NS5:\;\;\;&(x^0,x^1, x^2, x^3, x^4, x^5)\cr
NS5':\;\;&(x^0,x^1, x^2, x^3, x^8, x^9)\cr
D4:\;\;\;\;&(x^0, x^1, x^2, x^3, x^6)\cr
D6:\;\;\;\;&(x^0, x^1, x^2, x^3, x^7, x^8, x^9)\cr
D6':\;\;\;&(x^0, x^1, x^2, x^3, x^4, x^5, x^7)\cr
}}
One can check that any combination of two or more of these
branes preserves four or eight supercharges. The Lorentz
symmetry is broken in the presence of the branes to
\eqn\lorun{SO(9,1)\to SO(3,1)_{0123}\times SO(2)_{45}\times SO(2)_{89}}
(assuming that all branes that are pointlike in $(x^4, x^5)$ are placed
at the same point in the $(4,5)$ plane, and similarly for $(x^8, x^9)$).
To study situations with non-trivial $3+1$ dimensional physics
(in $(x^0, x^1, x^2, x^3)$) some of the branes must be made finite.

\lref\hanawit{A. Hanany and E. Witten, hep-th/9611230, Nucl. Phys.
{\bf B492} (1997) 152.}%
\lref\egk{S. Elitzur, A. Giveon and D. Kutasov, hep-th/9702014, 
Phys. Lett. {\bf B400} (1997) 269;
S. Elitzur, A. Giveon, D. Kutasov, E. Rabinovici and A. Schwimmer, 
hep-th/9704104, Nucl. Phys. {\bf B505} (1997) 202.}%

Consider a configuration \egk\ of $N_c$ $D4$-branes stretched between
an $NS5$-brane and an $NS5'$-brane separated by a distance $L$
along the $x^6$ direction (fig. 1).
At distances greater than $L$,
the five dimensional theory on the $D4$-branes reduces to a four
dimensional theory with $N=1$ supersymmetry. The boundary conditions
provided by the fivebranes imply that out of all the massless degrees
of freedom on the fourbranes (described by $4-4$ strings), only the
$4d$ $N=1$ vector multiplets for $G=U(N_c)$ survive. To decouple the
gauge dynamics from the complications of string theory one takes the
limit $g_s\to 0$, $L/l_s\to 0$, with the four dimensional
gauge coupling $g^2=g_sl_s/L$ held fixed \gkrev.

\vskip 1cm
\centerline{\epsfxsize=100mm\epsfbox{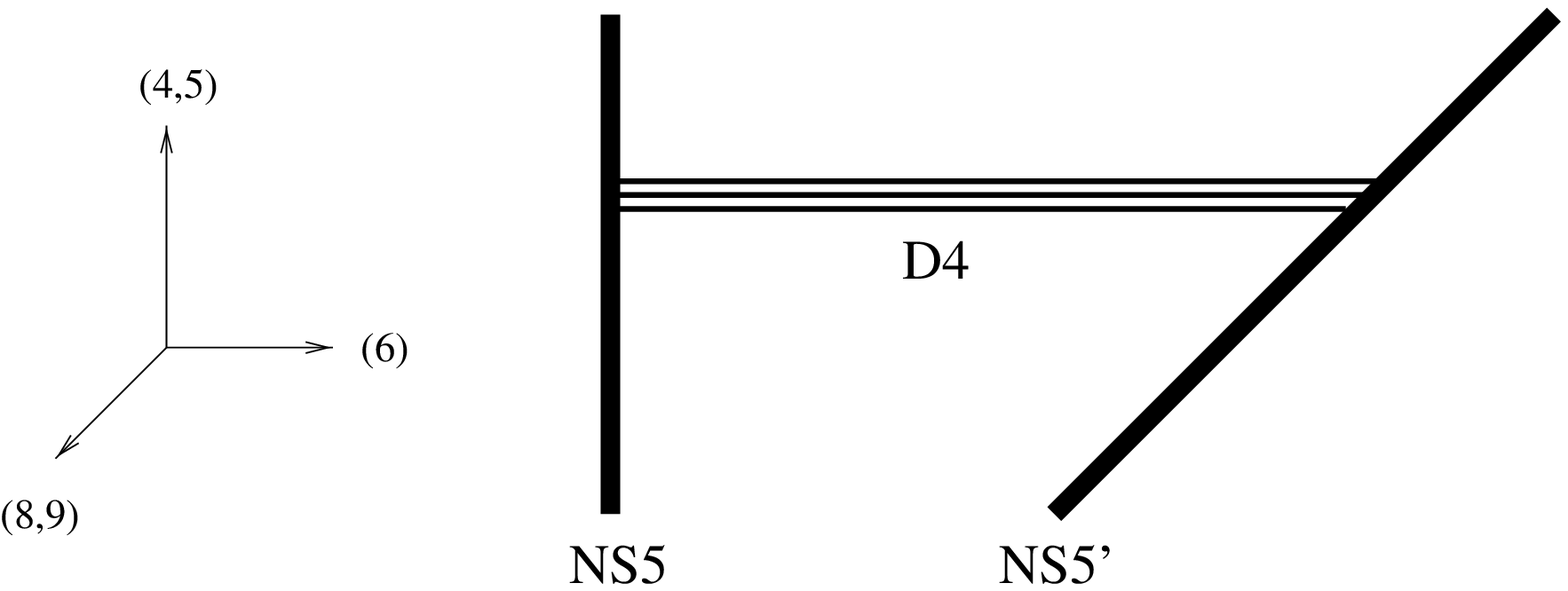}}
\centerline{Fig.\ 1}
\vskip .5cm

One can add matter in the fundamental representation
of $G$ in one of the two ways illustrated in fig. 2 (which, as
explained in \hanawit, are related).
One is to place $N_f$ $D6$-branes between the fivebranes
(fig. 2a). The $4-6$ strings give $N_f$ fundamentals of $U(N_c)$,
$Q^i$, $\tilde Q_i$, whose mass
is proportional to the separation between the sixbranes and the
fourbranes in $(x^4, x^5)$. Alternatively, one can add to the
configuration $N_f$ $D4$-branes stretching from the $NS5$-brane
to infinity (fig. 2b). In this case, the $N_f$ fundamentals of
$U(N_c)$ should arise from $4-4$ strings stretched between the
two kinds of fourbranes. Their mass is the separation between the
fourbranes in $(x^4, x^5)$.

\lref\brohan{J.H. Brodie and A. Hanany, hep-th/974043,
Nucl. Phys. {\bf B506} (1997) 157.}%

\vskip 1cm
\centerline{\epsfxsize=120mm\epsfbox{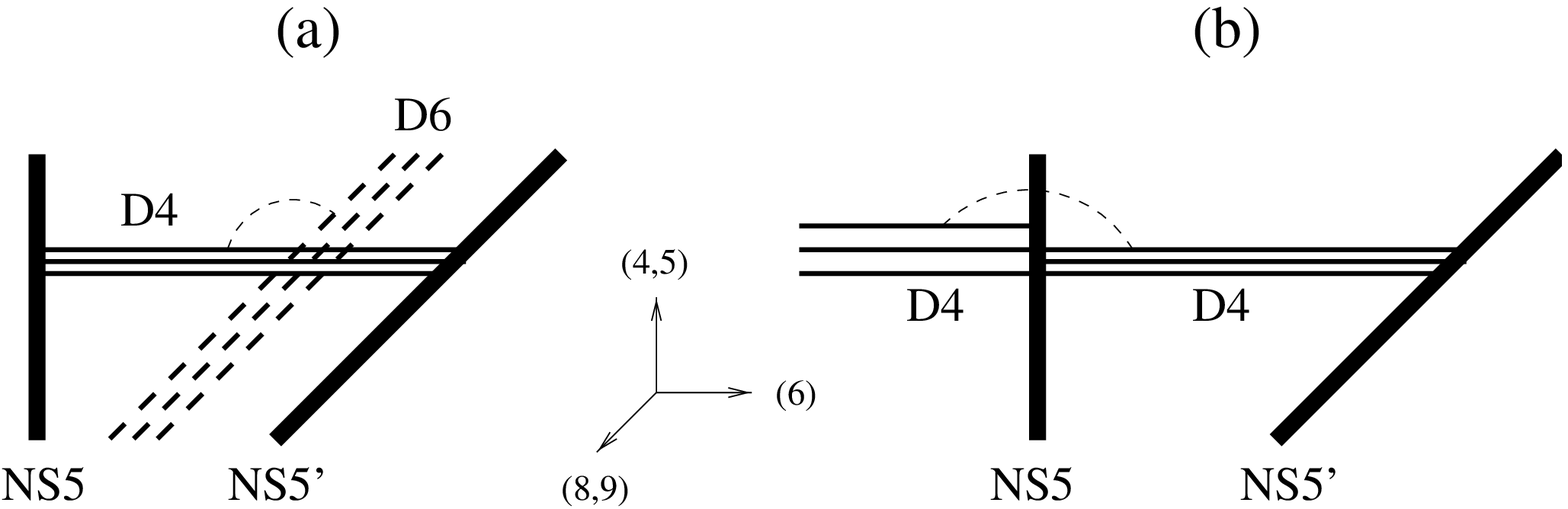}}
\centerline{Fig.\ 2}
\vskip .5cm

In the configuration of fig. 2a, the low energy physics should
be independent of the locations of the $D6$-branes along the
interval between the fivebranes. In \brohan\ it was pointed out
that a particularly natural location for the sixbranes is at the
same value of $(x^4, x^5, x^6)$ as the $NS5'$-brane. As is clear
from \nsd, in this case the $NS5'$-brane is embedded in the
$D6$-branes; it divides them into two disconnected parts (fig. 3).

\vskip 1cm
\centerline{\epsfxsize=65mm\epsfbox{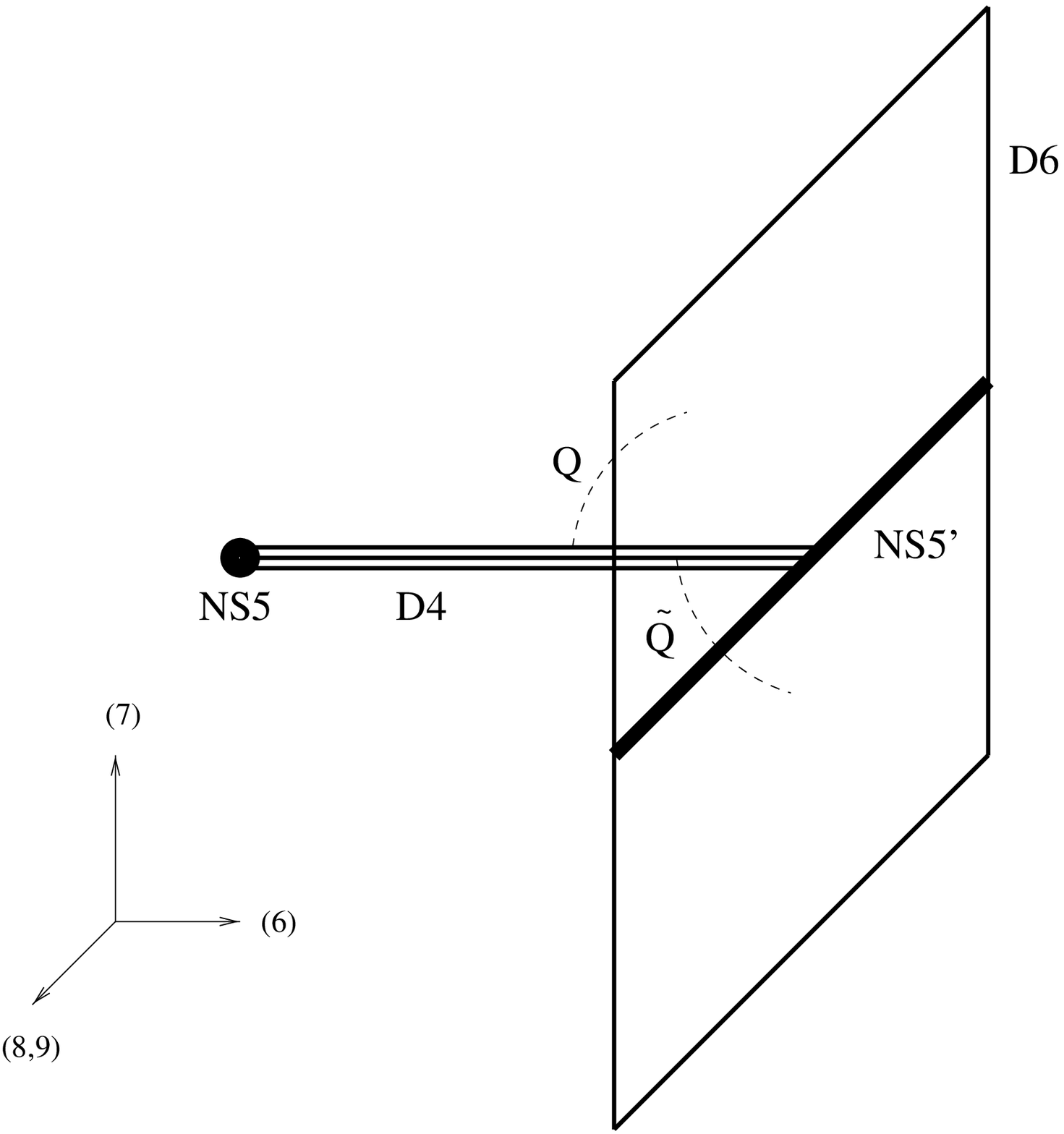}}
\centerline{Fig.\ 3}
\vskip .5cm

Consequently, the configuration has two separate $U(N_f)$ symmetries,
acting on the two semi-infinite sixbranes, and one may attempt to
interpret them as the $U(N_f)\times U(N_f)$ global symmetry of $N=1$
SQCD, under which $Q$ and $\tilde Q$ transform as $(N_F,1)$ and
$(1,\overline N_F)$, respectively.
This would imply that $4-6$ strings connecting the
$D4$-branes to the upper part of the $D6$-branes give rise at low
energies to the $N_f$ chiral multiplets in the fundamental of $U(N_c)$,
$Q^i$, while those that connect the fourbranes to the lower part of
the sixbranes give the $N_f$ chiral superfields in the anti-fundamental
of $U(N_c)$, $\tilde Q_i$, as indicated in fig. 3.

\lref\ahana{ A. Hanany and K. Hori, hep-th/9707192, Nucl. Phys.
{\bf B513} (1998) 119.}%
\lref\lll{K. Landsteiner, E. Lopez and D. Lowe, hep-th/9801002,
JHEP {\bf 9802} (1998) 007.}%
\lref\egkt{S. Elitzur, A. Giveon, D. Kutasov and D. Tsabar,
hep-th/9801020, Nucl. Phys. {\bf B524} (1998) 251.}%
\lref\bhkl{I. Brunner, A. Hanany, A. Karch and D. Lust,
hep-th/9801017, Nucl. Phys. {\bf B528} (1998) 197.}%

Despite the fact that the two groups of sixbranes are independent,
none of them can be removed from the configuration. From the
string theory point of view the reason is that this would lead to
violation of the RR charge that couples to $D6$-branes.
In the low energy gauge theory
the inconsistency is seen as a non-vanishing chiral anomaly.
However, the same basic mechanism for getting chiral matter
can be used to produce chiral spectra
in lower dimensions \ahana\ or in the presence of orientifolds
\refs{\lll,\egkt,\bhkl}.

The low-lying excitations of the brane configurations discussed
above can be divided into two classes: those that
are bound to one of the fivebranes, and those that are not. In this
paper we will analyze the properties of the first class of
excitations. It includes the following:

\item{(a)} $4-6$ strings connecting $N_c$ $D4$-branes to $N_F$
$D6'$-branes, all of which end on an $NS5$-brane\foot{The
configuration of fig. 4a can be obtained from that of fig. 3
by exchanging $(x^4, x^5)\leftrightarrow (x^8, x^9)$ and
removing some branes.} (fig. 4a).
The prediction is that  they give rise to a chiral spectrum: a
chiral superfield $Q$ in the $(N_c,N_F)$ of $U(N_c)\times U(N_F)$.
\item{(b)} $4-4$ strings connecting fourbranes ending on an
$NS5$-brane from opposite sides (fig. 4b). They should give rise
to chiral superfields, $Q$ in the $(N_c,N_F)$ and $\tilde Q$
in the $(\overline N_c,\overline N_F)$, or hypermultiplets $(Q, \tilde Q)$.
\item{(c)} $4-6$ strings connecting $D4$-branes ending on $NS5$-branes
to $D6$-branes intersecting the fivebranes (fig. 4c).
They should also give rise to hypermultiplets $(Q,\tilde Q)$.

\noindent
We will verify the predictions (a), (b), (c) below.
Excitations which belong to the second class and which we will
not analyze include the $4-4$ strings that give rise to the
$N=1$ vector superfield in fig. 1, and the $4-6$ strings that
give rise to the fundamental chiral superfields $Q^i$,
$\tilde Q_i$ in fig. 2a. The former are complicated to study,
since their wavefunction is spread throughout the interval
between the fivebranes and is influenced by them only via the
boundary conditions they provide. The latter can be studied
using standard $D$-brane techniques, at least when the intersection
of the $D4$ and $D6$-branes is far enough from the edges of the
fourbrane (see section 4). When that intersection lies on the
$NS5$-brane (fig. 4c), one can study it using the techniques of
this paper, and we will discuss this case below.

\vskip 1cm
\centerline{\epsfxsize=70mm\epsfbox{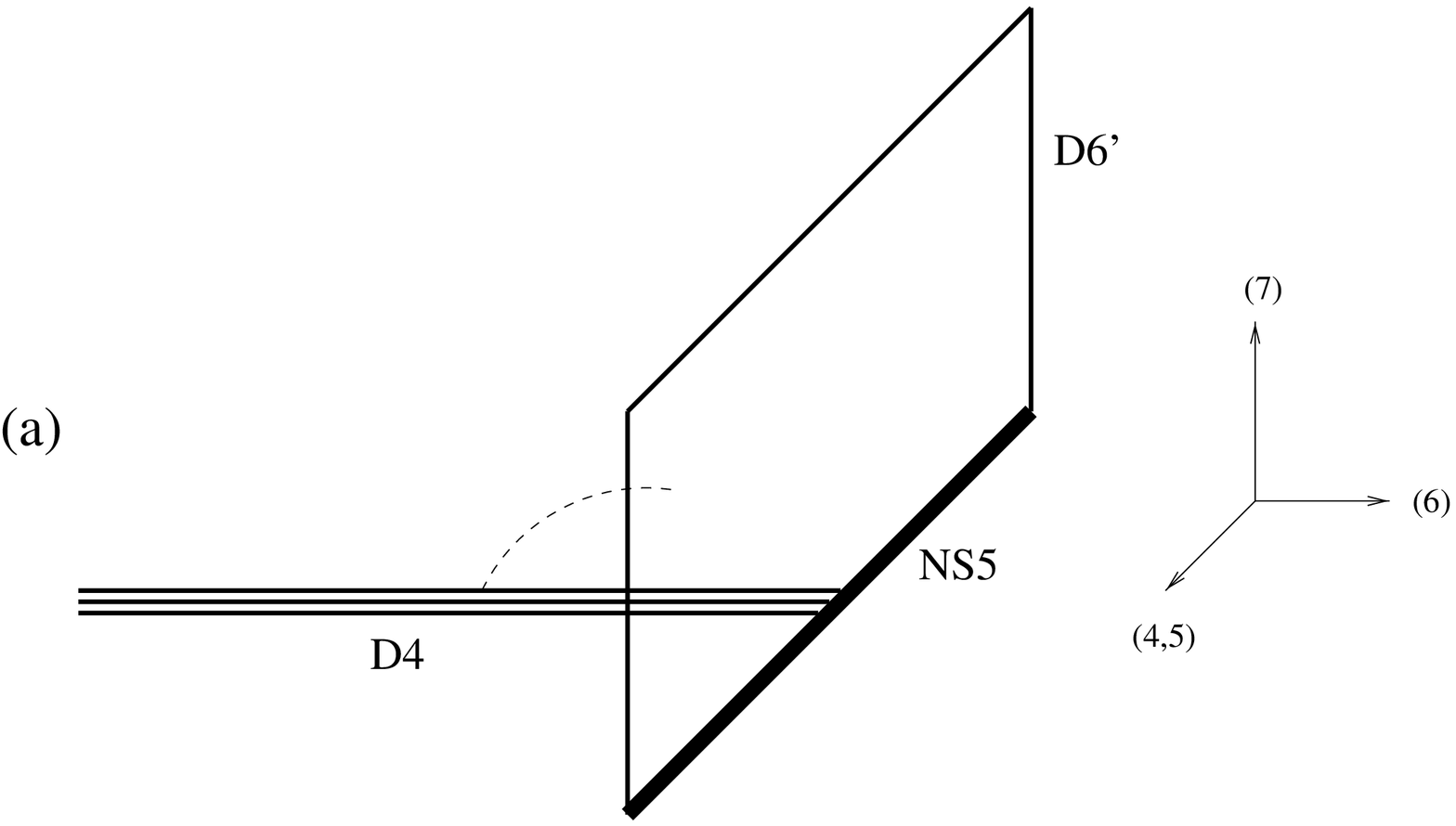}}
\vskip 0.5cm
\centerline{\epsfxsize=70mm\epsfbox{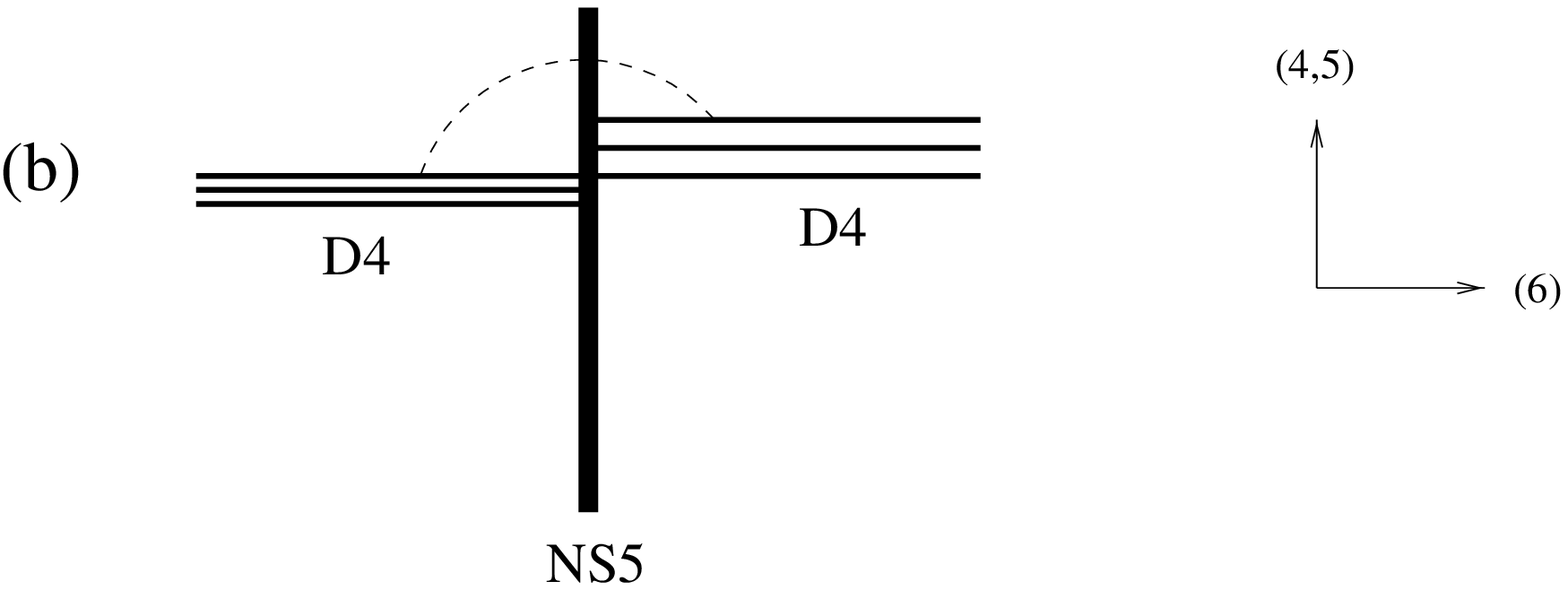}}
\vskip 0.5cm
\centerline{\epsfxsize=70mm\epsfbox{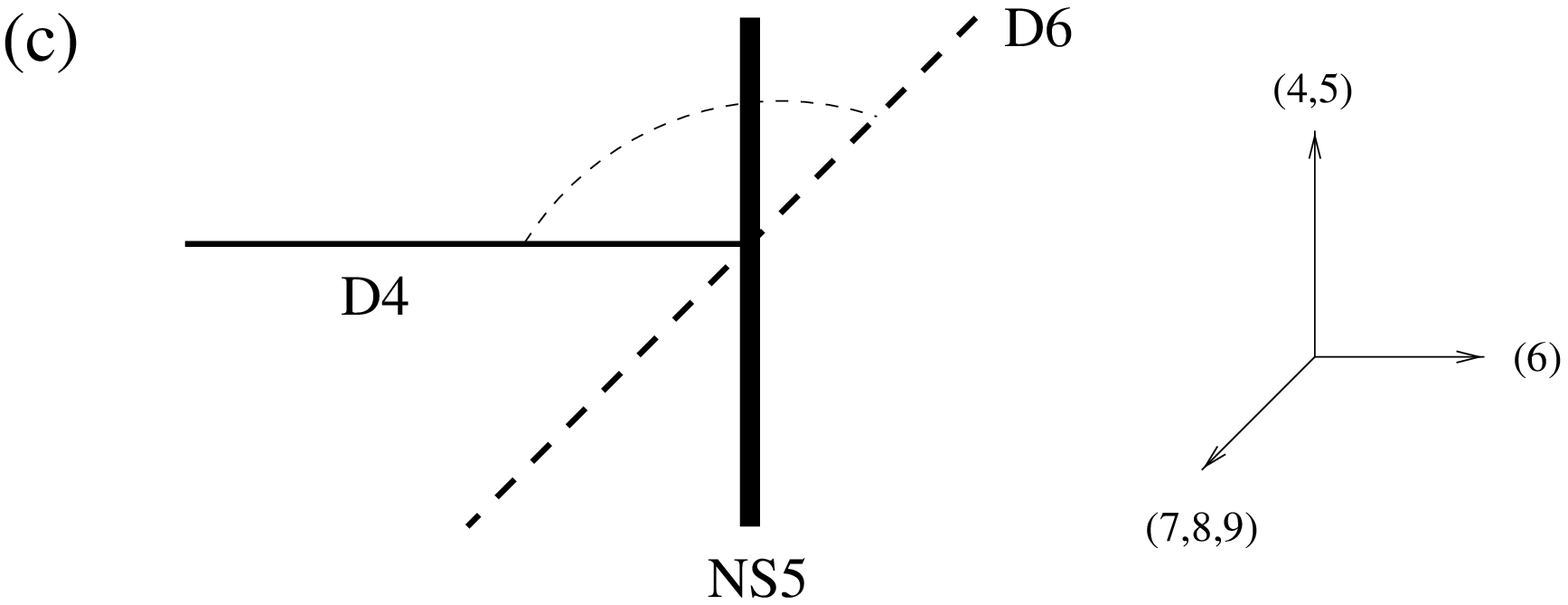}}
\centerline{Fig.\ 4}
\vskip .5cm

\newsec{The near-horizon geometry of $NS5$-branes and holography}

The background fields around a stack of $k$ parallel $NS5$-branes
are \chs:
\eqn\backfld{\eqalign{
&e^{2(\Phi-\Phi_0)}=1+\sum_{j=1}^k{l_s^2\over |\vec x-\vec x_j|^2}\cr
&G_{IJ}=e^{2(\Phi-\Phi_0)}\delta_{IJ}\cr
&G_{\mu\nu}=\eta_{\mu\nu}\cr
&H_{IJK}=-\epsilon_{IJKL}\partial^L\Phi\cr
}}
$I,J,K,L=6,7,8,9$ label the directions transverse to the
fivebranes (see \nsd). $\mu,\nu=0,1,\cdots, 5$ are the
directions along the brane. $\{\vec x_j\}$ are the locations
of the fivebranes in $\vec x =(x^6,\cdots, x^9)$. $H$ is the field
strength of the NS-NS $B$-field; $G$, $\Phi$ are the metric
and dilaton, respectively.

The background \backfld\ interpolates between flat
ten dimensional spacetime far from the fivebranes,
and a near-horizon region, in which the $1$ on the
right hand side of the first line of \backfld\ can
be neglected (fig. 5). This near-horizon region is
an asymptotically linear dilaton solution. E.g. if
the fivebranes are coincident, $\vec x_j=0$, the
near-horizon solution is
\eqn\chssol{\eqalign{
&e^{2(\Phi-\Phi_0)}={kl_s^2\over |\vec x|^2}\cr
&G_{IJ}=e^{2(\Phi-\Phi_0)}\delta_{IJ}\cr
&G_{\mu\nu}=\eta_{\mu\nu}\cr
&H_{IJK}=-\epsilon_{IJKM}\partial^M\Phi\cr
}}

\vskip 1cm
\centerline{\epsfxsize=80mm\epsfbox{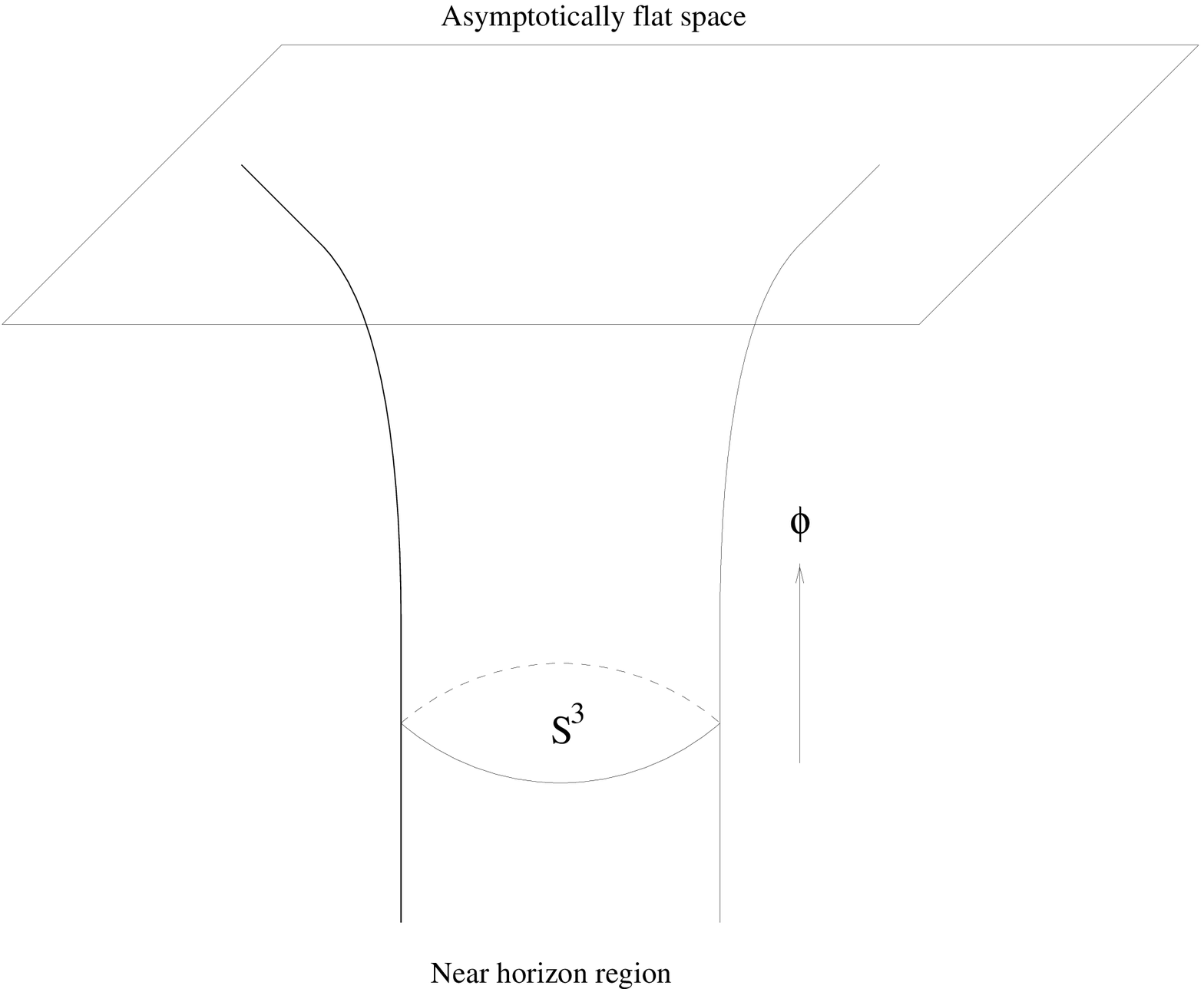}}
\centerline{Fig.\ 5}
\vskip .5cm

String propagation in the near-horizon
geometry \chssol\ can be described by an exact worldsheet
Conformal Field Theory (CFT) \chs.
The target space is
\eqn\throatg{\IR^{5,1}\times\IR_\phi\times SU(2)}
$\IR_\phi$ corresponds to the radial direction $r=|\vec x|$:
\eqn\phph{\eqalign{
&\phi={1\over Q}\log{|\vec x|^2\over kl_s^2}\cr
&\Phi=-{Q\over2}\phi\cr
}}
where we set $\Phi_0=0$ by rescaling $\vec x$.
$Q$ is related to the number of fivebranes via
\eqn\Qk{Q=\sqrt{2\over k}}
The CFT describing the three-sphere at constant $|\vec x|$
in \chssol\ is an $SU(2)$ WZW model at level $k$. The $SU(2)$
group element $g$ is related to the coordinates on the three-sphere
via (\eg):
\eqn\sutwo{g(\vec x)={1\over |\vec x|} \left[-x^6 1+
i(x^8 \sigma_1 + x^9 \sigma_2 +x^7 \sigma_3)\right]}
The $SO(4)\sim SU(2)_L\times SU(2)_R$ global symmetry
corresponding to rotations in the $\IR^4$ labeled by
$(x^6, x^7, x^8, x^9)$ acts on $g$ as $g\to h_L g h_R$,
where $h_{L(R)}\in SU(2)_{L(R)}$. Denoting the generators
of $SU(2)_L$ ($SU(2)_R$) by $J^a$ ($\bar J^a$), one finds
that $J^3-\bar J^3$ generates rotations in the $(x^6,x^7)$
plane, while $J^3+\bar J^3$ is the generator of rotations in
$(x^8, x^9)$.

Since we are dealing with the superstring, we are interested
in the $N=1$ superconformal $\sigma$-model on \throatg. Thus,
in addition to the bosonic coordinates $(x^\mu, \phi, g)$,
there are worldsheet fermions ($\psi^\mu, \chi^r, \chi^a$),
$a=1,2,3$, which are free (after a certain chiral rotation).
The worldsheet $N=1$ superconformal generators are:
\eqn\wssusy{\eqalign{
T(z)=&-{1\over2}(\partial x^\mu)^2-{1\over2}\psi^\mu\partial\psi_\mu
-{1\over2}(\partial\phi)^2-{Q\over2}\partial^2\phi-
{1\over2}\chi^r\partial\chi^r-{1\over k} J^aJ^a-
{1\over2}\chi^a\partial\chi^a\cr
G(z)=&i\psi_\mu\partial x^\mu +i\chi_r\partial \phi +iQ(\chi_aJ^a
-i \chi_1\chi_2 \chi_3+\partial \chi_r)\cr
}}
Here $J^a$ are the bosonic $SU(2)$ currents of level $k_B\equiv k-2$.
The total $SU(2)$ current algebra of level $k$ is generated by the
currents $J^a_{\rm total}=J^a+J^a_F$, where
$J^a_F=-(i/2)\epsilon^{abc}\chi_b\chi_c$ is the contribution of
the fermions. Note, in particular, that this construction only
makes sense for $k_B\ge 0$, \ie\ for two or more fivebranes. One
also imposes a chiral GSO projection $(-)^{F_L}=(-)^{F_R}=1$. The
GSO projected theory on the background \backfld, \throatg\ preserves
sixteen supercharges -- the $NS5$-brane is a half BPS object.

An interesting feature of the near-horizon geometry \chssol\ is that in
the vicinity of the fivebranes, $|\vec x|\to 0$, an infinite ``throat''
appears (for two or more fivebranes), corresponding to $\IR_\phi$ in
\throatg. In \abks\ it was proposed to interpret it in terms of holography.
String theory in the background \throatg\ was conjectured to be equivalent
to the theory on the fivebranes (LST).

The map between the ``bulk'' ($10d$) and ``boundary'' ($6d$) theories
is the following. On-shell observables in the bulk theory, such as vertex
operators in the background \throatg\ which correspond to non-normalizable
wavefunctions supported at the ``boundary,'' $\phi\to\infty$ on $\IR_\phi$,
are mapped to off-shell observables in the $6d$ fivebrane theory.
Non-normalizable vertex operators on \throatg\ depend among other things
on the six dimensional momentum $k^\mu$, which is interpreted as
off-shell momentum in the $6d$ LST.

Normalizable eigenstates of the Hamiltonian on \throatg\ correspond to
on-shell states in LST. One way to compute the spectrum of these states
is to study correlation functions of non-normalizable vertex operators
and look for singularities as a function of $k_\mu$. Poles at $k_\mu^2=
-M_i^2$ signal the presence of normalizable states with that mass in the
spectrum of the theory.

The worldsheet theory \chssol-\wssusy\ is singular. The string
coupling $g_s\sim\exp(-Q\phi/2)$ diverges as $\phi\to-\infty$
(\ie\ as one approaches the fivebranes \phph). Therefore, the
weakly coupled ten dimensional description is only useful for
studying those aspects of LST that can be analyzed at large
positive $\phi$. This includes identifying a large set of observables
(such as the aforementioned non-normalizable vertex operators),
and their transformation properties under the symmetries of the
theory (\eg\ $6d$ super-Poincare symmetry). Normalizable states
are difficult to analyze, since their wave-functions tend to
be supported in the strongly coupled region $\phi\to-\infty$.
Equivalently, correlation functions of non-normalizable operators
in linear dilaton vacua are typically not computable without
specifying the cutoff at large negative $\phi$.

If one is interested in studying the theory on $k>1$ coincident
fivebranes, one must face this strong coupling problem. However,
to make contact with section 2 we are in fact mainly interested
in the case where the fivebranes are separated. In that case one
might hope that the strong coupling region will be absent, \eg\
because the CHS solution with the throat only makes sense when
there are at least two coincident fivebranes. Indeed, one can show
that this is the case.

\lref\OV{H. Ooguri and C. Vafa, hep-th/9511164, \np{463}{1996}{55}.}%
\lref\sfetsos{K. Sfetsos, hep-th/9811167, JHEP {\bf 0001} (1999) 015;
hep-th/9903201.}%
\lref\gkp{A. Giveon, D. Kutasov and O. Pelc, hep-th/9907178,
JHEP {\bf 9910} (1999) 035.}%
\lref\gk{A. Giveon and D. Kutasov, hep-th/9909110, JHEP {\bf 9910}
(1999) 034;  hep-th/9911039, JHEP {\bf 0001} (2000) 023.}%
\lref\efrw{S. Elitzur, A. Forge and E. Rabinovici, 
Nucl. Phys. {\bf B359} (1991) 581;
G. Mandal, A.M. Sengupta and S.R. Wadia,
Mod. Phys. Lett. {\bf A6} (1991) 1685;
E. Witten, Phys. Rev. {\bf D44} (1991) 314.}%

If, for example, we distribute the $k$ fivebranes at equal
distances around a circle in the $(x^6, x^7)$ plane, which
breaks the global symmetry
\eqn\ggglll{SO(5,1)\times SO(4)\to SO(5,1)\times SO(2)\times Z_k}
the background \throatg\ changes as follows. Decompose
$SU(2)\to U(1)\times SU(2)/U(1)$. The $U(1)$ is the CSA
generator $J^3_{\rm total}$, which can be bosonized as
\eqn\jtbos{J^3_{\rm total}=2i\sqrt{k\over2}\partial Y}
where $Y$ is a scalar field normalized such that
$\langle Y(z) Y(w)\rangle=-{1\over4}\log(z-w)$, a normalization
that will be convenient later; it differs by a factor of
four from that of \refs{\gkp,\gk}.
In the original CHS solution, $(\phi, Y)$ describe an
infinitely long cylinder with a string coupling that varies
along the cylinder, diverging at one end and going to
zero at the other. Separating the fivebranes replaces
it \refs{\OV, \sfetsos, \gkp, \gk} with a semi-infinite cigar \efrw\
$SL(2)/U(1)$ in which the strong coupling region is absent,
or equivalently with $N=2$ Liouville, in which the
strong coupling region is suppressed by a superpotential
which goes like $\exp[-(\hat\phi+i\hat Y)/Q]$,
where $(\hat\phi, \hat Y)$ are the superfields whose bosonic
components are $(\phi, Y)$. For reasons explained
in \gk\ it is more convenient to use the coset (cigar) description
in this case. The full background replacing \chssol\ is
\eqn\backgr{ \IR^{5,1}\times {SL(2,\IR)_k\over U(1)}\times
{SU(2)_k\over U(1)}}
The global $SO(2)$ symmetry in \ggglll\ corresponds to translations
in $Y$ (\ie\ rotations of the cigar around its axis). The $Z_k$ charge
in \ggglll\ is the winding number around the cigar, a $U(1)$ which is
broken down to $Z_k$, since winding number can slip off the tip
of the cigar. It should be noted that the product in \backgr\ is not
direct, since the GSO projection relates the different factors. In
the GSO projected theory, the winding number around the cigar can
be fractional ($\in Z/k$). The fractional part of the winding number
is the $Z_k$ charge mentioned above.

\lref\teschner{J. Teschner,
hep-th/9712256, Nucl. Phys. {\bf B546} (1999) 390;
hep-th/9712258, Nucl. Phys. {\bf B546} (1999) 369;
hep-th/9906215, Nucl. Phys. {\bf B571} (2000) 555;
V.A. Fateev, A.B. Zamolodchikov and Al.B.
Zamolodchikov, unpublished.}%

Since the background \backgr\ can be made arbitrarily weakly
coupled\foot{by decreasing the value of the string coupling
at the tip of the cigar. This value is related to the radius
of the circle on which the fivebranes lie \gk.}, one can use
it to compute correlation functions in LST in a weak coupling
regime in its moduli space of vacua \gk. This is done by
constructing BRST invariant observables on \backgr\ and computing
their worldsheet correlation functions, using the results of
\teschner. For a detailed analysis we refer the reader to
\gk; here we mention a few facts that will play a role below.

\lref\seipol{N. Seiberg, Prog. Theor. Phys. Suppl. {\bf 102} (1990) 319;
J. Polchinski, Strings '90, Published in College Station Workshop, 1990.}

Consider the (NS,NS) sector of the theory\foot{The other
sectors can be reached by applying the spacetime supercharges
\gkp.}. The observables are primaries of the $N=1$ superconformal
algebra \wssusy\ with scaling dimension $(h,\bar h)=({1\over2},
{1\over2})$. The $\IR^{5,1}$ and $SU(2)/U(1)$ are well known SCFT's.
The former is a free field theory; the latter, an $N=2$ minimal
model with $c=3-(6/k)$. The $SL(2)/U(1)$ SCFT is $N=2$ superconformal
as well; it has central charge $c=3+(6/k)$.
The $N=2$ primaries $V_{j;m,\bar m}$ have scaling dimensions
\eqn\dimV{(h,\bar h)={1\over k}\left(m^2-j(j+1), \bar m^2-j(j+1)\right)}
and
\eqn\agmm{(m,\bar m)=\frac12(p+wk,p-wk)}
where $p,w\in Z$ are momentum and winding around the cigar, respectively.
As mentioned before, in the GSO projected theory \backgr\ one finds
\eqn\agwz{w\in {1\over k}Z}
while the momentum $p$ is still integer.

Unitarity and non-normalizability limit the range of $j$ to
\eqn\limj{j\in \IR~, \qquad-{1\over2}<j<{k-1\over 2}}
There are also delta-function normalizable operators with $j=-{1\over2}
+is$, $s\in \IR$. These do not give rise to off-shell observables in the
theory, but rather should be thought of as producing a continuum of states
above a gap in LST\foot{The relation between states and operators in the
cigar CFT is subtle and very similar to that in Liouville theory, described
in \seipol.}. All other observables (in addition to $V_{j;m,\bar m}$)
can be obtained by acting with $N=2$ superconformal generators on these
primaries.

\lref\gkunpub{A. Giveon and D. Kutasov, unpublished.}

By analyzing correlation functions of $V_{j;m,\bar m}$, one finds \gk\ that
poles in correlators correspond to discrete representations\foot{In some
cases one also finds poles corresponding to $m=j$ \gkunpub.} of $SL(2)$
\eqn\disrepp{|m|=j+n,\;\;|\bar m|=j+\bar n;\;\;n, \bar n=1,2,3,\cdots}
This leads to a discrete spectrum of states in LST, which exhibits
Hagedorn growth at high energy.

\newsec{Some properties of $D$-branes}

In this section we review some properties of $D$-branes in flat space
and on $S^3$, in preparation for our discussion of $D$-branes in the
CHS geometry \throatg, and its regularized version \backgr.

\subsec{4-6 strings in flat space}

\lref\polcha{J. Polchinski, S. Chaudhuri and C.V. Johnson, hep-th/9602052.}%
\lref\hash{A. Hashimoto, hep-th/9608127, Nucl. Phys. {\bf B496} (1997) 243.}%

Later, we will analyze $4-6$ strings connecting $D4$-branes and
$D6'$-branes, both of which end on a stack of $NS5$-branes (fig. 4a).
We start by reviewing the simpler case of intersecting infinite $D4$
and $D6'$-branes in flat space \refs{\polcha, \hash}.

Consider an open string, one of whose ends is on a $D4$-brane.
The other end may be either on the same brane or on
another brane. We want to study the emission of a $4-6'$ string
from the $D4$ boundary of this string. After the emission, this
boundary of the emitting string will lie on a $D6'$-brane (fig. 6).

\vskip 1cm
\centerline{\epsfxsize=40mm\epsfbox{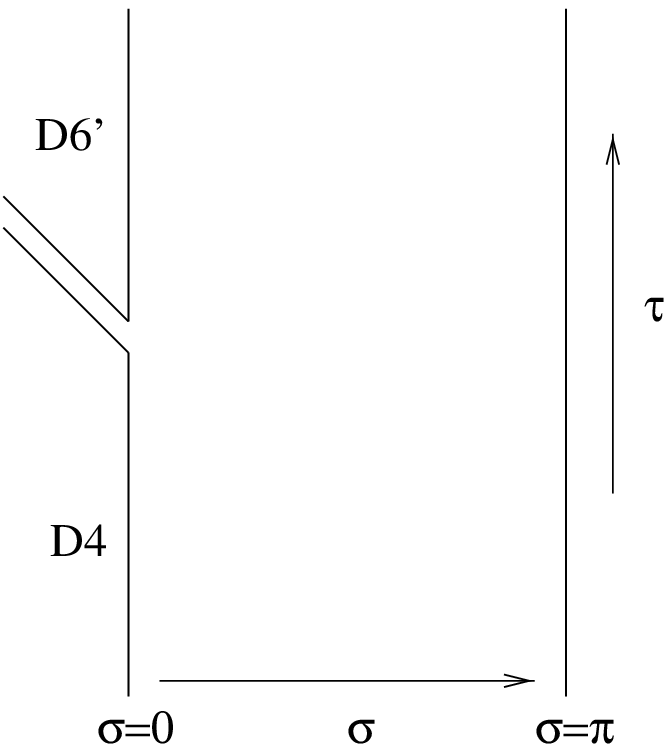}}
\centerline{Fig.\ 6}
\vskip .5cm

The worldsheet of an open string is a strip, or equivalently
the upper half plane. On the upper half plane, worldsheet
time evolves radially. Equal time surfaces are half circles
around the origin, with early times corresponding to small
circles. The boundary of the upper half plane (the real line
${\rm Im} z=0$) corresponds to the ends of the string and is
divided into two parts: $z>0$ or $\sigma=0$ in fig. 6,
which lies on the $D4$-brane, and $z<0$ or $\sigma=\pi$.

The emission of an open $4-6'$ string from the boundary
$\sigma=0$ is described by an insertion of a vertex operator
$V$ at some point $z>0$ on the boundary. The boundary now
splits into three components:
(i) $z'<0$, the spectator boundary at $\sigma=\pi$, (ii)
$0<z'<z$, which lies on the fourbrane, and (iii) $z'>z$,
which is on the sixbrane (see fig. 7).
The boundary conditions for, say, the coordinate $x^4$ are Dirichlet,
$\partial x^4(z') +\overline\partial x^4(z') =0$, in region (ii), and
Neumann, $\partial x^4(z') - \overline \partial x^4(z')=0$, in region
(iii).

\lref\difran{P. Di Francesco, P. Mathieu and D. Senechal,
{\it Conformal Field Theory}, NY Springer 1997.}%

\vskip 1cm
\centerline{\epsfxsize=100mm\epsfbox{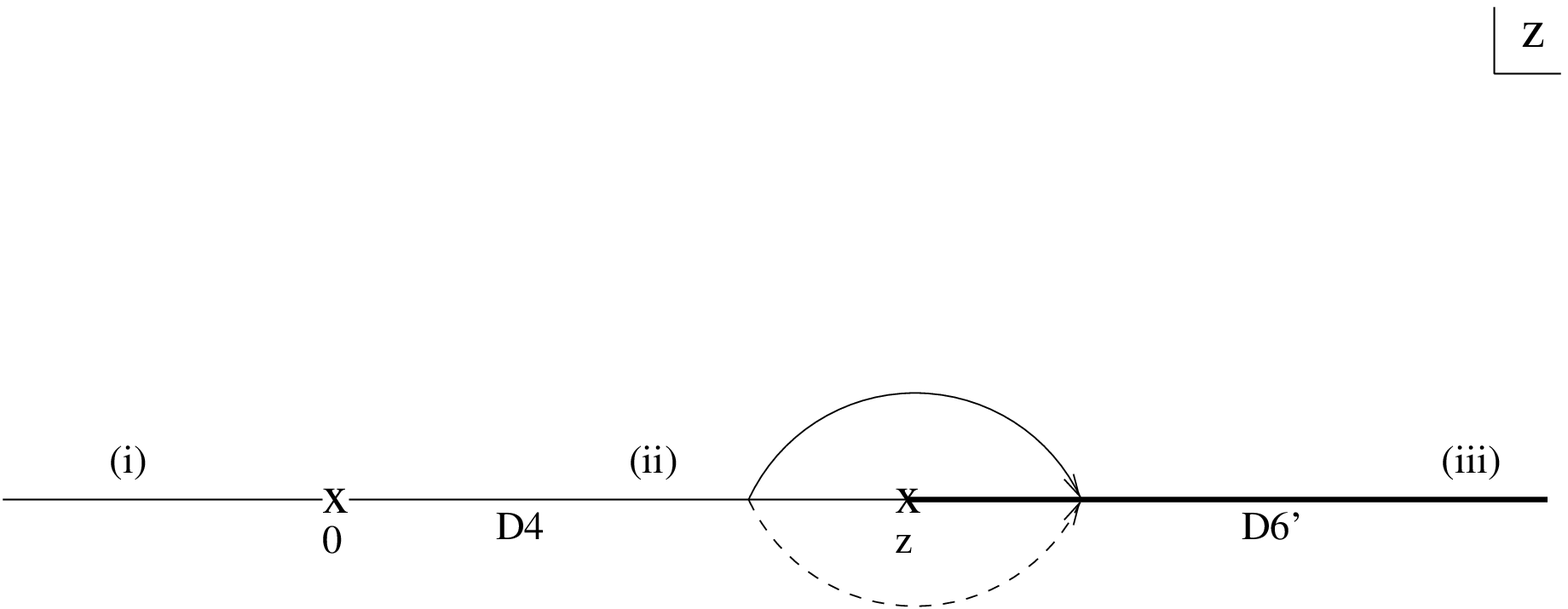}}
\centerline{Fig.\ 7}
\vskip .5cm

The vertex operator $V(z)$ changes the boundary conditions of
$x^4$ from Neumann to Dirichlet. If one inserts the holomorphic
current $\partial x^4$ into the worldsheet, and moves it along a
small upper semicircle around $z$ (see fig. 7), from $z'<z$ to
$z'>z$ (the operator $\overline\partial x^4$ is necessarily
transported along the mirror image lower semicircle between
these two points), the relative sign of the two operators has
to flip. This means that the
operator product $V(z) \partial x^4(z')$ has to have a square root
branch cut in $z-z'$. Operators with such a cut are twist operators,
familiar from orbifold CFT (see \eg\ \difran). The lowest dimension
operator of this type, $\sigma$, has dimension $1/16$.

The same arguments apply to all the coordinates for which the $4-6'$
string has Dirichlet-Neumann boundary conditions; hence, the operator
$V$ contains a twist field for $(x^4, x^5, x^6, x^7)$, $\sigma_{4567}$,
with total dimension $1/4$. We will be mainly interested in emission
of spacetime bosons, in which case $V$ belongs to the $NS$ sector of
the worldsheet CFT. Therefore, it has to be local with respect to
$G =i\sum \psi_a\partial x^a$, the worldsheet superconformal generator.
Since $G$ has a square root branch cut with respect to $x^{4,5,6,7}$,
it must also have a cut with respect to $\psi_{4,5,6,7}$. This implies
that $V$ has to contain the spin field $S_{4567}$ for these
worldsheet fermions. This operator also has dimension $1/4$.

The directions $x^{0,1,2,3}$ are common to the $D4$ and $D6'$-branes,
and are treated as in standard open string theory with Neumann
boundary conditions. The remaining coordinates $(x^8, x^9)$ have
Dirichlet-Dirichlet boundary conditions. If the $D4$-brane is, say,
at $x^8=0$, while the $D6$-brane is at $x^8=a$ (and both at $x^9=0$),
then similarly to the discussion above, a bulk operator of the form
$\exp(i k x^8)$ when moved around the above mentioned semi-circle from
a point on the $D4$ to a point on the $D6'$, has to pick a factor of
$\exp(ik a/2 )$, together with another such contribution from the mirror
path of the lower semicircle. This means that the operator product
$V(z) \exp(ik x^8(z'))$ has a branch cut of the form $(z-z')^{ka/2\pi}$
and the product with the corresponding right-moving operator has a cut
with the opposite sign. The appropriate boundary operator is
$\exp(i(a/\pi)( x^8_L(z)-x^8_R(z)))$. This operator generates
``winding number'' $a/\pi$ along $x^8$ -- in agreement with the
geometrical picture of a $4-6'$ string stretched a distance $a$
along $x^8$. Its dimension is $a^2/2\pi^2$.

Collecting all the factors, we get the following vertex operator
describing the (bosonic) ground state of a $4-6'$ string stretched
between a $D4$-brane at $(x^8, x^9)=(0,0)$ and a $D6'$-brane at
$(x^8, x^9)=(a,b)$,
\eqn\flverop{V=e^{-\varphi} \sigma_{4567} S_{4567}
e^{{i\over\pi}(a (x^8_L-x^8_R)+b (x^9_L-x^9_R))} e^{ik_\mu x^\mu}}
where $k_\mu$ ($\mu=0,1,2,3$) is the $4d$ spacetime momentum.
$\varphi$ is the bosonized superconformal ghost. The vertex operator
\flverop\ is written in the $-1$ picture; one can check that the
coefficient of $e^{-\varphi}$ in \flverop\ is an $N=1$ superconformal
primary, which is a necessary condition for its BRST invariance. The
requirement that it has worldsheet dimension $1/2$, which is also
necessary for BRST invariance, implies that the mass squared of the
ground state of the $4-6'$ string is $-k_\mu^2 = {1\over\pi^2}
(a^2+b^2)$, as one would expect (we work in a convention $\alpha'=1/2$,
in which the scalars $x$ are canonically normalized on the boundary
and the tension of the fundamental string is $T=1/\pi$).

In particular, when the $D6'$-brane intersects the $D4$-brane, \ie\ when
$a=b=0$, this mass vanishes. The vertex operator \flverop\
describes a particle which transforms as a scalar under $3+1$ dimensional
Lorentz rotations.
The spin field $S_{4567}$ has $4$ components, half of which are
projected out by the GSO projection, so \flverop\ actually describes
two real scalar particles. For the applications described in section 2
it is sometimes useful to consider not one but a stack of $N$ $D4$-branes.
In that case, the two scalars \flverop\ transform in the fundamental
representation (${\bf N}$) of the $U(N)$ gauge symmetry on the fourbranes.

In addition to the $4-6'$ strings described above there are also $6'-4$
strings which have similar properties but transform in the ${\bf \overline N}$
of $U(N)$. Altogether we have two complex scalars, $Q$ in the ${\bf N}$ and
$\tilde Q$ in the ${\bf \overline N}$ of $U(N)$. The system of $D4$ and
$D6'$-branes preserves eight supercharges ($N=2$ SUSY in the four dimensions
$(0123)$) and acting with the supercharges on \flverop\ completes
a hypermultiplet transforming in the fundamental representation
of $U(N)$.

\subsec{$D$-branes on the $SU(2)$ group manifold}

\lref\pss{G. Pradisi, A. Sagnotti and Y.S. Stanev, hep-th/9503207,
Phys. Lett. {\bf B354} (1995) 279; hep-th/9506014,
Phys. Lett. {\bf B356} (1995) 230; hep-th/9603097,
Phys. Lett. {\bf B381} (1996) 97.}%
\lref\klim{C. Klimcik and P. Severa, hep-th/9609112,
Nucl. Phys. {\bf B488} (1997) 653.}%
\lref\katoo{M. Kato and T. Okada, hep-th/9612148,
Nucl. Phys. {\bf B499} (1997) 583.}%
\lref\biasta{ M. Bianchi and Y.S. Stanev, hep-th/9711069,
Nucl. Phys. {\bf B523} (1998) 193.}%
\lref\alekschom{A. Alekseev and V. Schomerus, hep-th/9812193,
Phys. Rev. {\bf D60} (1999) 061901.}%
\lref\stanold{S. Stanciu, hep-th/9901122, JHEP {\bf 9909} (1999) 028.}%
\lref\gaw{K. Gawedzki, hep-th/9904145.}%
\lref\bfs{L. Birke, J. Fuchs and C. Schweigert, hep-th/9905038.}%
\lref\gcpleb{H. Garcia-Compean and J.F. Plebanski, hep-th/9907183.}%
\lref\bppz{R.E. Behrend, P.A. Pearce, V.B. Petkova and J.-B. Zuber,
hep-th/9908036.}%
\lref\arsold{A. Alekseev, A. Recknagel and V. Schomerus, hep-th/9908040,
JHEP {\bf 9909} (1999) 023.}%
\lref\fffs{G. Felder, J. Frohlich, J. Fuchs and C. Schweigert,
hep-th/9909030.}%
\lref\stanciu{See, for example,
S. Stanciu, hep-th/9909163, JHEP {\bf 0001} (2000) 025,
and references therein.}%
\lref\fofs{J.M. Figueroa-O'Farrill and S. Stanciu, hep-th/0001199.}%
\lref\bdsnew{C. Bachas, M. Douglas and C. Schweigert, hep-th/0003037.}%
\lref\arsnew{A. Alekseev, A. Recknagel and V. Schomerus, hep-th/0003187.}%
\lref\cardy{J.L. Cardy, Nucl. Phys. {\bf B324} (1989) 581.}%

We next turn to some facts regarding $D$-branes on a group manifold $G$,
focusing on the case $G=SU(2)$ ($D$-branes on group manifolds have been
studied, for instance, in
\refs{\pss, \klim, \katoo, \biasta, \alekschom, \stanold, \gaw, \bfs,
\gcpleb, \bppz, \arsold, \fffs, \stanciu, \fofs, \bdsnew, \arsnew}).
In the absence of $D$-branes, the WZW
model has an affine $G_L\times G_R$ symmetry. If $g(z,\overline z)$ is
a map from the worldsheet to the group $G$, the symmetry acts on it as:
\eqn\bulksym{g \to h_L(z) g h_R(\overline z )}
If the worldsheet has a boundary, there is a relation between left-moving
and right-moving modes, and the $G_L\times G_R$ symmetry is broken. One can
still preserve some diagonal symmetry $G$, say the symmetry
\eqn\bcsym{g\to h g h^{-1}}
corresponding to $h_L=h_R^{-1}=h$ in \bulksym. The presence of this symmetry
constrains the boundary conditions that can be placed on $g$.
Allowing  $g({\rm boundary})=f$  for some $f\in G$ we must also allow
$g({\rm boundary})=hfh^{-1}$ for every $h\in G$. This means that
$g$ on the boundary takes value in the conjugacy class containing $f$
\alekschom. For $G=SU(2)$, conjugacy classes are parametrized by a single
angle $\theta$, $0\leq \theta \leq \pi$, corresponding to the choice
$f=\exp(i\theta \sigma_3)$.
Thinking of $SU(2)$ as the group of three dimensional
rotations, the conjugacy class $C_{\theta}$ is the set of all rotations
of angle $2\theta$ about any axis. A boundary condition which
preserves \bcsym\ is then
\eqn\symbc{g({\rm boundary})\in C_{\theta}}
Since  $h_L$ is generated by the currents $J^a$ while $h_R$ is generated
by $\overline J^a$, the invariance of the
boundary condition \symbc\ under \bcsym\
implies that the currents satisfy \stanciu:
\eqn\curbc{J^a=\overline J^a;\,\,\,a=1,2,3}
on the boundary.
Not any value of $\theta$ in \symbc\  gives rise to a
consistent model \refs{\klim, \alekschom, \gaw}.
Recall that for a general group $G$, the level $k$ WZW action has the form
\eqn\wzterm{S= \int_\Sigma d^2z L^{sm}+\int_B L^{WZ}}
where $L^{sm}={k\over{4\pi}} Tr(\partial g^{-1}
\overline{\partial} g)$ is the sigma model part of the action
and $L^{WZ}={k\over {4\pi}}\omega^{(3)}$ is the Wess-Zumino term.
Here $\Sigma $ is the worldsheet Riemann surface,
$\omega^{(3)}={1\over 3}Tr((g^{-1}dg)^3)$ is a closed three-form,
and $B$ is a three dimensional manifold whose boundary is the
worldsheet. When the worldsheet $\Sigma$ has itself boundaries,
it cannot be the boundary of a three dimensional manifold, since
a boundary cannot have boundary. To define the WZ term in \wzterm\
for this case, one should fill the holes in the worldsheet by adding
discs, and extend the mapping from the worldsheet into the group
manifold to these discs. One further demands that the whole disc
$D$ is mapped into a region (which we will also refer to as $D$)
inside the conjugacy class in which the corresponding boundary lies.
$B$ will then be defined as a three-manifold bounded by the union
$\Sigma\bigcup D$, which now has no boundaries.

The resulting action should preserve the symmetry \bulksym\
in the bulk of the worldsheet which tends to \bcsym\  on
the boundary. Take $C$, the conjugacy class containing $D$,
to be the class of a fixed group element $f$, {\it i.e.}
\eqn\conj{C=\{hfh^{-1}|h\in G\}}
Let $\delta_L g$ be an infinitesimal variation of $g(z,\bar z)$ in
the bulk such that $\overline {\partial} (\delta_L g g^{-1})=0$
and $\delta_R g$ a variation for which $\partial(g^{-1}\delta_R g)=0$.
On the boundary of $\Sigma$, where $g\in C$ is parametrized as $g=hfh^{-1}$,
$\delta_L g g^{-1} = -g^{-1}\delta_R g= \delta h h^{-1}$. By the above
symmetry the variation $(\delta_L + \delta_R) S$ should vanish. In the
interior of the disc we should allow an arbitrary variation $\delta h$
since the location of the auxiliary disc inside $C$ has no physical
significance and cannot influence the action. Using the identity
\eqn\delome{\delta \omega^{(3)}=dTr((g^{-1}\delta g)(g^{-1}dg)^2)}
one gets
\eqn\vars{(\delta_L + \delta_R) S= {k\over {4\pi}}\left[\int_D
Tr((g^{-1}\delta g)(g^{-1}dg)^2)+\int_{\partial \Sigma} A\right]}
where $A$ is the one-form
\eqn\a{A=Tr [(h^{-1}\delta h)(f^{-1}h^{-1}dhf-fh^{-1}dhf^{-1})]}
The first term in \vars\ comes from the change in the region of
integration of $L^{WZ}$ resulting from the variation of $D$. The
second term is a boundary correction to the symmetry \bulksym.
Clearly the right hand side of \vars\  is not zero for a non-trivial
mapping of $\partial \Sigma $ and of $D$ into $C$.

To fix the symmetry we have to modify the action by adding to it
an integral over the disc $D$ of some two form $\omega^{(2)}$,
defined on $C$, such that its variation will cancel \vars. The
proper action has now the form
\eqn\pact{S= \int_\Sigma d^2z L^{sm}+\int_B L^{WZ} + {k\over {4\pi}}
\int_D \omega^{(2)}}
The variation of this action is
\eqn\varms{(\delta_L + \delta_R) S={k\over {4\pi}}
\left[\int_D Tr((g^{-1}\delta g)(g^{-1}dg)^2)-
 \int_{\partial D} A +\int_D \delta \omega^{(2)}\right]}
where we used the fact that $\partial \Sigma $ and $\partial D$
are identical curves with opposite orientations.
The vanishing of \varms\ fixes $\delta \omega^{(2)}$ as
\eqn\delometwo{\delta \omega^{(2)}=dA- Tr((g^{-1}\delta g)
(g^{-1}dg)^2)}
Expressing the r.h.s. of \delometwo\ in terms of the group
element $h$ parametrizing $C$ as in \conj\  one gets
$\delta \omega^{(2)}$ explicitly on $C$. Its solution is
\refs{\alekschom, \gaw}
\eqn\ometwo{\omega^{(2)}=Tr[(h^{-1}dh)f^{-1}(h^{-1}dh)f]}
Notice that (as implied by \delometwo) on the conjugacy
class $C$, $d\omega^{(2)}=\omega^{(3)}$.

The modified action \pact\ is independent, by construction,
of continuous deformations of $D$ inside $C$. However, in
general, the second homotopy of a conjugacy class is non-trivial.
If we compare then the value of the action for $D$ and $D'$, two
different choices of embedding the disc in $C$ with the same boundary,
$D'$ may not be a continuous deformation of $D$ in $C$. In that
case the above analysis does not imply that the two ways to evaluate
the action \pact\ agree. Since there is no natural way to choose
between the two embeddings, \pact\ is not yet a well defined action.
In particular, for $G=SU(2)$ the conjugacy classes $C$ have the
topology of $S^2$, the two-sphere generated by all possible axes
of rotation by a fixed angle in three dimensions. One may then
choose $D$ and $D'$ such that their union covers the whole of $S^2$.
In that case the difference between the action $S_D$, the value of
\pact\ with embedding $D$, and $S_{D'}$ with embedding $D'$ is
\eqn\dels{\Delta S= {k\over {4\pi}}
\left[\int_B \omega^{(3)}+\int_C\omega^{(2)}\right]}
where $B$ is the three-volume in $SU(2)$ bounded by the two-sphere
$C$ (fig. 8).

\vskip 1cm
\centerline{\epsfxsize=45mm\epsfbox{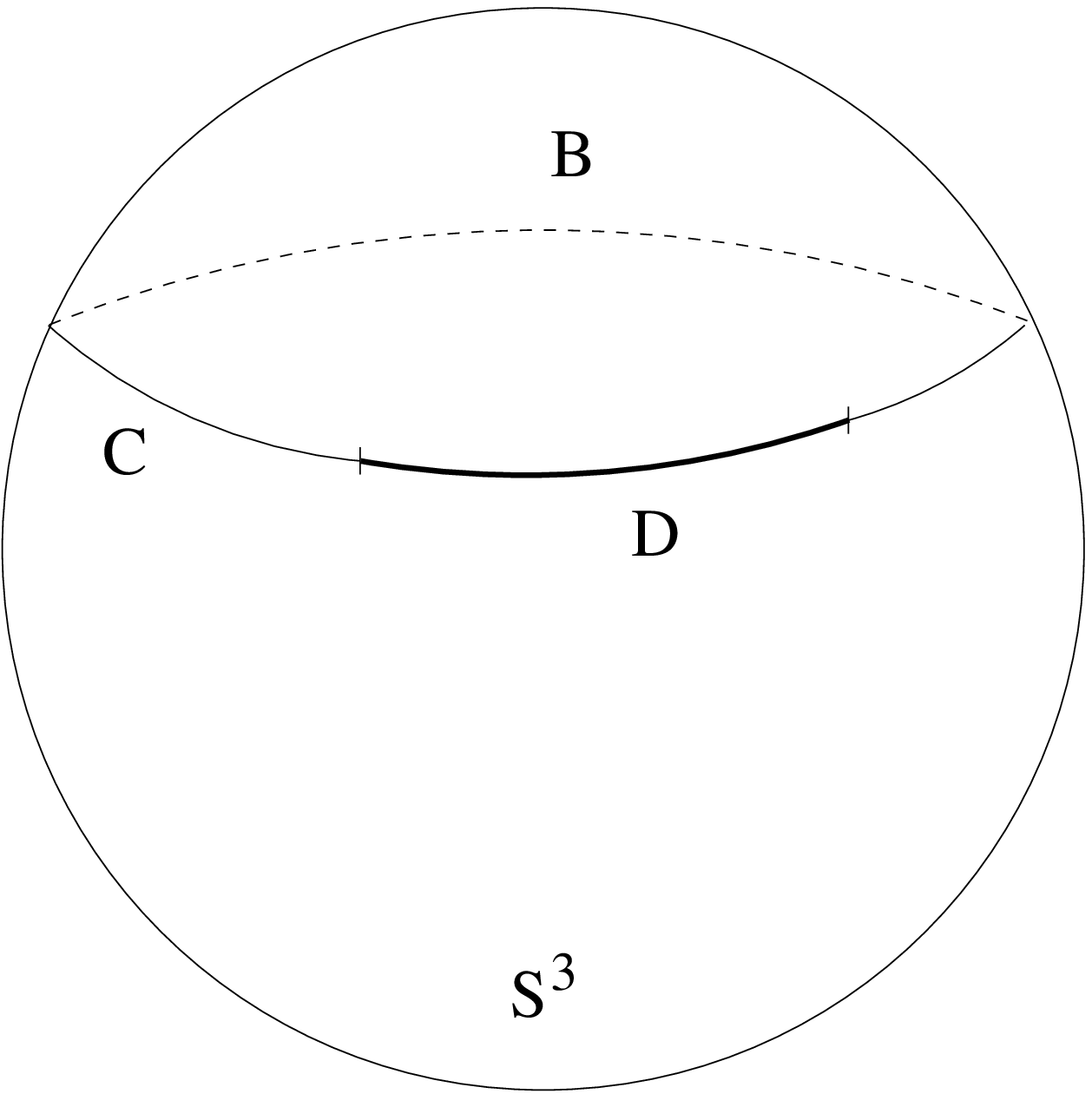}}
\centerline{Fig.\ 8}
\vskip .5cm

For the case of $SU(2)$, which has the topology of $S^3$,
the form $\omega^{(3)}$ is $\omega^{(3)}=4\Omega^{(3)}$ where
$\Omega^{(3)}$ is the volume form on the unit three-sphere.
For $C$ in \conj\ corresponding to
$f=\exp(i \theta \sigma_3)$, the first term in \dels\ is
\eqn\vol{\int_B \omega^{(3)}=8\pi(\theta - {1\over 2}\sin(2\theta))}
As to the two-form $\omega^{(2)}$, from eq. \ometwo\ we see
that $\omega^{(2)}[h]=\omega^{(2)}[qh]$ for any fixed element
$q \in SU(2)$. This implies that this form is proportional to
$\Omega^{(2)}$, the volume form of the unit two-sphere. The
expression \ometwo\ for $h$ near the identity gives then
for the conjugacy class $C_{\theta}$,
\eqn\twovol{\omega^{(2)}=\sin(2\theta)\, \Omega^{(2)}}
This gives for the change in the action for two topologically
different embeddings in \dels\
\eqn\deltheta{\Delta S= 2k\theta}
Although this is non-zero, the quantum theory is
still well defined if $\Delta S$ is an integral multiple of
$2\pi$. We find then that the possible conjugacy classes on which
a boundary state can live are quantized, the corresponding $\theta$
must satisfy
\eqn\qtheta{\theta=2\pi {j\over k}}
with $j$ integer or half integer satisfying $0 \le j \le {k\over 2}$.

Thus, there are $k+1$ different boundary conditions preserving
the symmetry \bcsym\ which one can impose on the $SU(2)_k$ group
manifold.  This matches nicely with the algebraic analysis of Cardy
\cardy, who found that in a general rational CFT on a worldsheet
with boundary, to each primary field of the chiral algebra there
corresponds a  boundary state preserving the diagonal chiral algebra,
the analog of \bcsym. In the case of the $SU(2)$ WZW model, the chiral
algebra is affine $SU(2)_k$, and there are $k+1$ different primary
fields corresponding to representations of (integer or half integer)
spin $0\le j \le {k\over 2}$. It is natural, \refs{\alekschom, \gaw},
to identify each of the $k+1$  different geometric boundary conditions
corresponding to $j$ in \qtheta\  with the algebraic boundary state
corresponding to the same $j$.

Since all of these boundary states preserve \bcsym, the open strings
stretching between them should have well defined transformation properties
under the diagonal chiral algebra. According to \cardy, the open strings
stretched between a boundary state corresponding to a primary field $j$
and another boundary state corresponding to primary field $j'$, belong to
the representations of the chiral algebra which appear in the fusion of $j$
and $j'$.

We have chosen the boundary conditions \symbc\ such that the particular
diagonal subgroup $G$  of $G\times G$ defined in eq. \bcsym\ will survive
them. This is of course not a unique choice. One can act on these boundary
conditions with any element of the $G\times G$ symmetry group to get
equivalent boundary conditions which preserve a different diagonal subgroup.
Thus we can multiply the conjugacy class $C_{\theta}$  in eq. \symbc\ from
the right (which is the same as multiplying from the left) by any group
element $f$ to get modified boundary conditions
\eqn\mbc{g({\rm boundary})\in C_{\theta} f}
These boundary conditions also preserve a diagonal subgroup, since the set
$C_{\theta} f$  satisfies
\eqn\conjf{ C_{\theta} f = h (C_{\theta} f) f^{-1} h^{-1} f}
for any $h\in G$. Therefore, the boundary conditions \mbc\ preserve
the diagonal subgroup of $G_L\times G_R$ defined by
$h_R= f^{-1}h_L^{-1}f$.
In terms of the infinitesimal generators of $G_L\times G_R$, \ie\ the left
and right handed currents, the invariance of \mbc\ under this subgroup
implies for the corresponding boundary state the condition
\eqn\mcurbc{J^a = (Ad_{f^{-1}}\overline J)^a}
which modifies \curbc\ by conjugating the right handed
currents by $f^{-1}$.

\newsec{$D$-branes in the near-horizon geometry of $NS5$-branes}

After assembling the necessary tools, we are now ready to study the
physics of the configurations of fig. 4.

\subsec{$D4$ and $D6'$-branes ending on $NS5$-branes}

We start with the configuration of fig. 4a. A stack of $N_c$
$D4$-branes ends from the left, {\it i.e.} from negative $x^6$,
on $k$ coincident $NS5$-branes. $N_F$ $D6'$-branes end on the
$NS5$-branes from above (positive $x^7$). From the point of view
of the geometry \backfld\ (fig. 5), the $D$-branes extend into
the CHS throat, as indicated in fig. 9a.

The $D4$-branes intersect the three-sphere at the point
$x^7=x^8=x^9=0$; the $D6'$-branes at $x^6=x^8=x^9=0$
(see fig. 9b). Thus,
\sutwo, they correspond to the boundary states $g|_{\rm boundary}
=1$ and $i\sigma_3$, respectively. The $D4$-branes are described by the
boundary state with $\theta=0$ and $f=1$ (see \symbc, \qtheta), while the
$D6'$-branes correspond to a transformed state \mbc, with $\theta=0$
and $f=\exp(i\pi\sigma_3/2)=i\sigma_3$.
The $SU(2)$ currents $J^a$ satisfy
the boundary conditions \curbc\ and \mcurbc\ for strings ending on
the $D4$ and $D6'$-branes, respectively. In order to preserve worldsheet
supersymmetry one has to impose analogous boundary conditions on the
fermions. For example, for a string ending on the $D4$-branes one
has
\eqn\ferbc{\chi^a=\overline \chi^a}
while for a boundary on a $D6'$-brane the $\chi^a$ satisfy an analog
of \mcurbc. Since, as is clear from fig. 9, both the fourbranes and
the sixbranes extend into the throat, the boundary conditions on $\phi$
are Neumann.

\vskip 1cm
\centerline{\epsfxsize=140mm\epsfbox{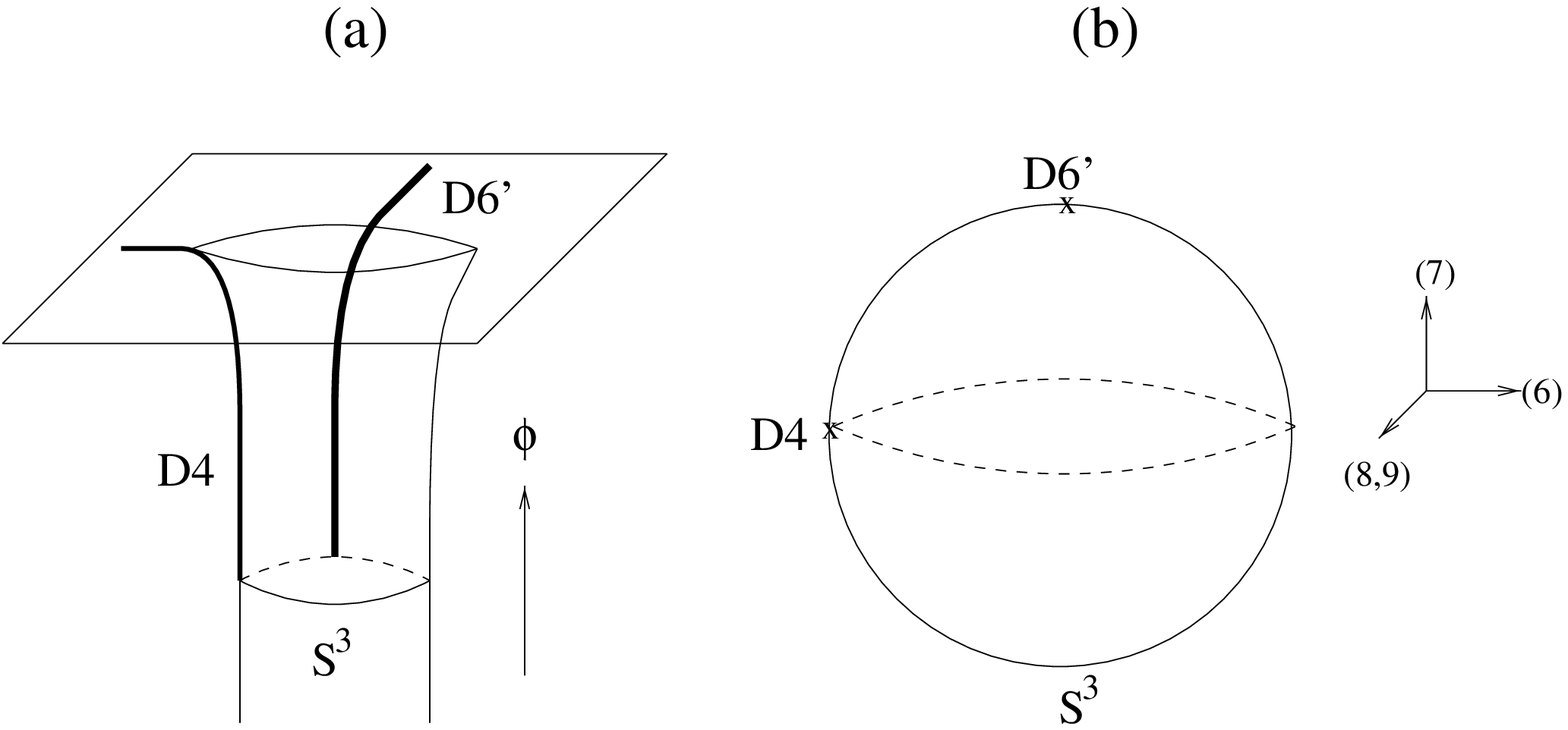}}
\centerline{Fig.\ 9}
\vskip .5cm

We would like to construct the vertex operator for emitting
the lowest lying $4-6'$ string in the geometry of fig. 9,
\ie\ generalize \flverop\ to the fivebrane near-horizon
geometry. Some parts of the discussion leading to \flverop\
are unchanged. In particular, the geometry is the same as there in
$(x^0, x^1, x^2, x^3, x^4, x^5)$. The presence of $\phi$ allows
also a contribution $\exp(\beta\phi)$ to the vertex operator. Thus,
the analog of \flverop\ for this case has the form
\eqn\ffflll{V=e^{-\varphi} \sigma_{45} S_{45}
e^{ik_\mu x^\mu}e^{\beta\phi}V_2}
where $V_2$ is the contribution of the $SU(2)$ group manifold
to the vertex operator, to which we turn next.

The $4-6'$ vertex operator $V_2$ changes the worldsheet boundary
conditions from $g=1$ to $g=f=\exp(i\alpha\sigma_3/2)$. The
$D6'$-brane corresponds to $\alpha=\pi$, but it is instructive
to discuss the general case, in which the angle between the $D4$
and $D6'$-branes is $\alpha/2$ (see fig. 10).

\vskip 1cm
\centerline{\epsfxsize=140mm\epsfbox{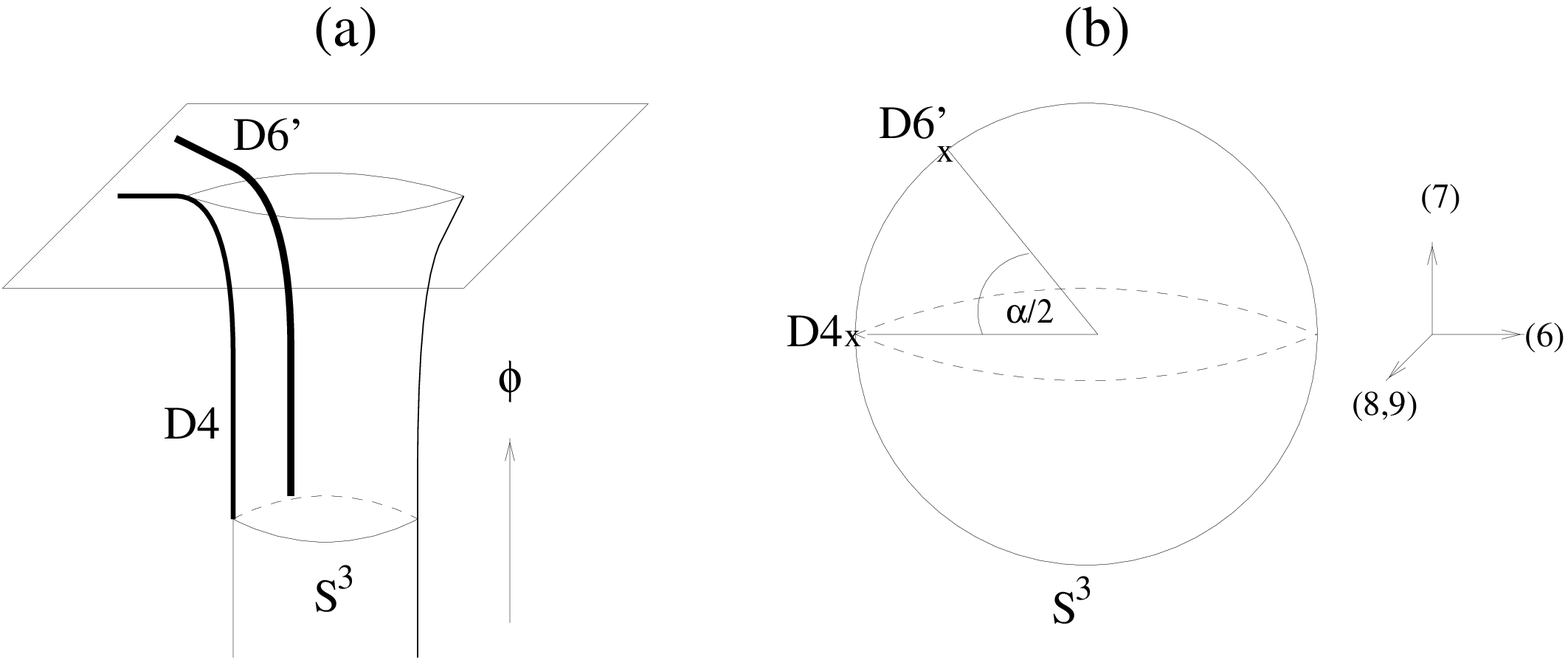}}
\centerline{Fig.\ 10}
\vskip .5cm

As discussed above,
$g=1$ corresponds to the boundary condition \curbc, \ferbc\ for the
$SU(2)$ currents and fermions, while $g=f$ gives rise to \mcurbc
\eqn\alphacurbcthree{\eqalign{&J^3=\overline J^3\cr
&\chi^3=\overline\chi^3\cr
&J^\pm =\exp(\mp i\alpha) \overline J^\pm \cr
&\chi^\pm =\exp(\mp i\alpha) \overline \chi^\pm\cr}}
Note that the symmetry generated by $J^3+\bar J^3$ is preserved
by both \curbc\ and \alphacurbcthree. As discussed after eq. \sutwo,
it corresponds to the rotation symmetry $SO(2)_{89}$ which is unbroken
by the brane configuration. As in section {\it 4.1}, we conclude that
$V_2$ must include a twist field $\sigma_\alpha$ with the following
locality properties w.r.t. the left-moving currents $J^a$ and fermions
$\chi^a$:
\eqn\sigj{\eqalign{
&\sigma_\alpha(z) J^3(z')\sim (z-z')^{m_1}O_1(z')\cr
&\sigma_\alpha(z) \chi^3(z')\sim (z-z')^{m_2}O_2(z')\cr
&\sigma_{\alpha}(z) J^\pm (z')\sim (z-z')^{\mp{\alpha \over2 \pi}
+n_1}O_3(z')\cr
&\sigma_{\alpha}(z) \chi^\pm (z')\sim
(z-z')^{\mp{\alpha \over {2 \pi}}+n_2}O_4(z')\cr}}
with $m_i,n_i\in Z$; $O_I$ on the r.h.s. are operators whose precise
form will not be specified here.
A similar twist holds for the right-movers.
Equation \sigj\ implies that $V_2$ belongs to a representation
of a twisted affine $SU(2)$, in which the currents $J^\pm$ have
fractional modes. This twisted algebra reads
\eqn\twi{\eqalign{
&[J^3_n, J^3_m]=n{k_B\over 2}\delta_{n,-m}\cr
&[J^3_n, J^\pm_{m \pm {\alpha\over {2\pi} }}]=
\pm J^\pm_{n+m \pm {\alpha\over {2\pi}}}\cr
&[J^+_{n+ {\alpha\over {2\pi}}},J^-_{m-{\alpha\over {2\pi}}}]=
(n+ {\alpha\over {2\pi}})k_B\delta_{n,-m} + 2 J^3_{n+m}\cr
}}
where $k_B=k-2$ is the level of the bosonic $SU(2)$ algebra (as
in section 3).
The algebra \twi\ can be mapped into the standard (untwisted)
affine Lie algebra by using spectral flow. If $J^a$ satisfy the
twisted algebra \twi, the generators $\tilde J$ defined by
\eqn\tildth{\eqalign{
&\tilde J^3_n=J^3_n +{k_B\over 2} {\alpha\over 2\pi}\delta_{n,o}\cr
&\tilde J^\pm_n=J^\pm_{n \pm {\alpha \over 2 \pi}}\cr
}}
satisfy the ordinary untwisted algebra. Thus one can use standard
facts about the representations of untwisted affine Lie algebra to
study the twisted one.

The modes of the energy-momentum tensor ($\tilde L_n$)
constructed from $\tilde J$ are related to those of the original
energy-momentum tensor ($L_n$) via
\eqn\enmo{\tilde L_n=L_n + {\alpha\over{2 \pi}}J^3_n
+{k_B\over 4} {\alpha ^2 \over {4 \pi^2}}\delta_{n,0}}
To understand the properties of the operator $\sigma_\alpha$,
consider first the case $\alpha=0$. $\sigma_0$ describes
an open string connecting two boundary states, both
corresponding to $j=0$ in eq. \qtheta. As discussed
in section {\it 4.2}, this open string should transform
in a representation of the diagonal $SU(2)$ contained in
the fusion of two $j=0$ representations, which
consists of only the spin $0$ representation. If $h$ is the
energy ($L_0$) and $m$ the $J^3$ charge, then the lowest energy
state in this sector has $h=m=0$ and all the excited states
satisfy
\eqn\ineq{h\geq |m|}
Turning on $\alpha$ continuously, the open string created by
$\sigma_\alpha$ remains in the spin 0 representation, now of
the affine algebra generated by the currents $\tilde J$
\tildth. Hence the corresponding states satisfy
\eqn\tilineq{\tilde h \geq |\tilde m|}
or, using \enmo,
\eqn\ineqtp{h\geq(\pm 1-{\alpha \over {2 \pi}})
\tilde m+{k_B\over 4}{\alpha^2\over 4\pi^2}}
where $\tilde m \in Z $. For $0\le \alpha <2\pi$,
the lowest energy state corresponds to $\tilde m =0$.
The dimension and charge of the corresponding operator are
\eqn\minen{\eqalign{
h=&{k_B\over 4} {\alpha^2\over 4\pi^2}\cr
m=&-{k_B\over 2} {\alpha\over 2 \pi}\cr}}
To construct this operator it is convenient to decompose the
$SU(2)_{k_B}$ bosonic CFT under $U(1) \times SU(2)/U(1)$,
where $U(1)$ represents the subgroup generated by $J^3$.
One can bosonize\foot{As in section {\it 4.1}, here and below
scalar fields are canonically normalized on the boundary;
\eg\ $\langle u(z) u(w)\rangle=-\log|z-w|$ for $z,w\in R$.}
$J^3$ as in \jtbos, $J^3= 2i\sqrt{k_B/2}\,\partial u$; $J^\pm $
are represented by $\exp(\pm 2i\sqrt{2/k_B}\, u)$ multiplied by
operators from the $SU(2)/U(1)$ coset. The operators
\eqn\jver{\sigma^B_{\alpha}=
\exp\left[-i \sqrt {k_B\over 2} ({\alpha\over {2\pi}}+n)u\right]}
with $n\in Z$, have the right locality properties \sigj\
with respect to the currents. The dimension and charge \minen\
correspond to those of the operator \jver\ with $n=0$. Note that
for $\pi\le\alpha <2\pi$, setting $n=-1$ in \jver\ would give a lower
dimension than that of $n=0$, however, the charge and dimension do
not satisfy the inequality \ineqtp\ in this case. Hence, these
operators are not in the spectrum.

The operator $\sigma_\alpha$ must also flip the boundary conditions
of the fermions $\chi^a$ from \ferbc\ to \alphacurbcthree. This
means that it must include a spin field for $\chi^\pm$.
Bosonizing the two fermions,
\eqn\boschipm{i\partial H=\chi^+\chi^-}
one finds that the operators with the right locality properties
w.r.t. $\chi^\pm$ are
\eqn\bosfer{\Sigma_n = e^{ i(n-{\alpha \over2\pi})H}}
where $n\in Z$.
For $0\leq \alpha <\pi$ the operator of lowest dimension out of these
is $\Sigma_0$ whose scaling dimension is
$h=\alpha^2/8\pi^2$ and $U(1)$ charge $-\alpha/2\pi$. Thus,
in this range of $\alpha$,
the lowest lying state in the sector twisted by $\alpha$ is
\eqn\lowsig{\sigma_\alpha =e^{-i{\alpha\over 2\pi}H}
e^{-i{\alpha\over2\pi}\sqrt{k_B\over2}u}}
Notice that  $\sigma_\alpha$ is a twist field for
$J_{\rm total}$, \ie\ one can rewrite it as
\eqn\tottwist{\sigma_\alpha=e^{-i\sqrt{k\over2}{\alpha\over2\pi}Y}}
where $Y$ is defined in \jtbos\ and is related to $u$ and $H$ via
\eqn\yuh{\sqrt{k\over2}Y=\sqrt{k_B\over2}u+H}
The operators \bosfer\ with $n\neq 0$ can be thought of as
``excited'' twist fields.
For $\pi<\alpha\leq 2\pi$ the operator
$\Sigma_1$ in \bosfer\ has lower dimension
then that of $\Sigma_0$. Then the excited twist operator
\eqn\exctwist{\sigma'_\alpha=e^{-i({\alpha \over2\pi}-1)H}
e^{-i\sqrt{k_B\over2}{\alpha\over2\pi}u}}
has the smallest dimension in the $\alpha$ twisted sector\foot{Applying
the GSO projection to \ffflll\ will pick up different spin fields
$S_{45}$ for the different twist fields \lowsig\ and \exctwist.}.
For $\alpha=\pi$, the supersymmetric case, the two operators in \lowsig\
and in  \exctwist\ are degenerate on the worldsheet.
Nevertheless, we show in Appendix B that the operator \exctwist\
creates from the vacuum excited $4-6'$ strings, even for $\alpha=\pi$.

To summarize, the lowest lying open string connecting the $D4$
and $D6'$-branes in the configuration of fig. 10 (which we will
refer to as a $4-6'_+$ string, since it connects the fourbrane
to a half-infinite $D6'$-brane at $x^7>0$) is described by the
vertex operator
\eqn\vfoursixpr{V_{4-6'}^+=e^{-\varphi} \sigma_{45} S_{45}
e^{i k_\mu x^\mu} e^{\beta \phi}
e^{-i{\alpha\over2\pi}\sqrt{k\over2}Y}}
The mass shell condition requires that
\eqn\mmaass{{1\over2}k_\mu^2-{1\over2}\beta(\beta+Q)+
{k\over4}\left({\alpha\over2\pi}\right)^2={1\over4}}
We see that the $4-6'_+$ string vertex \vfoursixpr\
describes an on-shell excitation in {\it five} non-compact
dimensions, $(x^0, x^1, x^2, x^3, \phi)$. This is
clearly the same phenomenon as that discussed for
closed strings in section 3. The extra non-compact
dimension is the infinite throat of \chs\ associated
with the radial direction. Non-normalizable
open string vertex operators such as \vfoursixpr\ with
$\beta >-Q/2$ correspond to {\it off-shell observables}
in the theory on the $D$-branes.

To find the spectrum of low lying on-shell states,
which is relevant for analyzing the physics of the brane
configurations described in section 2, one has to find
the normalizable states corresponding to $4-6'_+$ strings.
One way of doing that is to compute
the correlation functions of the non-normalizable operators
\vfoursixpr\ and extract the spectrum by analyzing their
analytic structure. As explained in section 3, to do this
within a weakly coupled string theory one has to regularize
the strong coupling region $\phi\to-\infty$. Following
\gk, we will do this by considering the geometry \backgr\
corresponding to fivebranes separated along a circle in
the $(x^6, x^7)$ plane\foot{One might wonder whether the
separation of the fivebranes introduces a finite mass shift
for $4-6'$ strings, when the fourbrane and sixbrane end on
different fivebranes. To see that this does not happen note
that, as discussed in \gk, the cigar geometry corresponds
to the limit $g_s, r_0/l_s\to 0$ with $r_0/(l_sg_s)$ fixed
and large ($r_0$ is the typical separation between fivebranes).
The mass shift for $4-6'$ strings associated with this separation
is $\sim r_0/l_s^2$; it goes to zero in string units in the above
limit.}.

The effect of this regularization on the vertex operator
\vfoursixpr\ is to change the wavefunction on the infinite
cylinder labeled by $(\phi, Y)$ to a wavefunction on the
semi-infinite cigar,
\eqn\cylcig{e^{\beta \phi}e^{-i{\alpha\over2\pi}\sqrt{k\over2}Y}
\to V_{jm}}
where
\eqn\mappar{\eqalign{j=&\beta\sqrt{k\over2}\cr
                      m=&-{k\over2}{\alpha\over2\pi}\cr
}}
$j$ can be thought of as momentum along the cigar, while $m$
is the momentum around the cigar (both the $D4$-branes and
$D6'$-branes are wrapping the cylinder, and in particular
the circle labeled by $Y$).

The vertex operator \vfoursixpr\ becomes in the geometry
\backgr\
\eqn\vvff{V_{4-6'}^+=e^{-\varphi}\sigma_{45}S_{45}
e^{i k_\mu x^\mu} V_{jm}}
with \mmaass
\eqn\mmmaaa{{k_\mu^2\over 2}+{m^2-j(j+1)\over k}={1\over4}}
To compute the spectrum of low-lying normalizable excitations
of the $4-6'_+$ string, one computes the two point function
of the operators \vvff\ on the disk. This amplitude exhibits
first order poles in $j$. Using the mass-shell condition
\mmmaaa, these can be interpreted as poles in $k_\mu^2$; they
correspond to on-shell particles in four dimensions, created from
the vacuum by the operator \vvff.

The calculation of this two point function is very similar to its
closed string analog \refs{\teschner,\gk}, and is described in
appendix A. The result
is, as in the closed string case, a series of poles corresponding
to the discrete representations of $SL(2)$, \disrepp. The lowest
lying state corresponds to $n=1$, \ie\ $j=|m|-1$.
Plugging this
together with \mappar\ into \mmmaaa\ we find that the mass of
the lowest lying normalizable state of the $4-6'_+$ string  is
\eqn\alphmass{M^2(\alpha)={1 \over 2}({\alpha\over\pi}-1)}
In particular, we find that as expected, for $\alpha=\pi$ the
lowest lying state is massless. The vertex operator which creates
this massless particle from the vacuum is
\eqn\agrec{V_{4-6'}^+(k_\mu^2=0)=
e^{-\varphi}\sigma_{45}S_{45}e^{i k_\mu x^\mu}
V_{{k\over 4}-1,-{k\over 4}}}
Note that the degeneracy is correct too. The spin field $S_{45}$
in \vvff\ takes apriori two possible values,
but the GSO projection picks one of them. Thus, \vvff\ describes
a single complex scalar field $Q$ which transforms in the $(N_c,
N_F)$ of the $U(N_c)\times U(N_F)$ symmetry on the $D$-branes,
as explained in section 2. Applying the spacetime supercharges
gives rise to a chiral superfield \foot{In particular, one can
check that \vvff, \agrec\ is annihilated by half of the spacetime
supercharges.} $Q$. The complex conjugate field $Q^\dagger$
arises from $4-6'_+$ strings with the opposite orientation
(a $6'_+ -4$ string). The field $\tilde Q$ (${\tilde Q}^\dagger$)
transforming in the $(\overline N_c, \overline N_F)$
representation that is needed for anomaly cancellation of $N=1$ SQCD
comes from $6'_- -4$ ($4-6'_-$) strings connecting the fourbranes to
sixbranes attached to the $NS5$-branes from below, as anticipated in \brohan.
These are described by vertex operators of the form \vfoursixpr,
\vvff, \agrec\ with $\alpha<0$ and with the $S_{45}^\dagger$ spin field
for the $4-6'_-$ operator $\tilde Q^\dagger$ (replacing
$S_{45}$ for the $4-6'_+$ operator $Q$).\foot{The reason
for the $S_{45}^\dagger$ is the
following. The $D6'_-$ and $D6'_+$-branes must have the same orientation.
They can be obtained by starting with parallel $D6'$ and $\bar D6'$-branes
(at $\alpha=0$). The $D6'_+$-brane is obtained by rotating the $D6'$-brane
by an angle ${\alpha\over2}={\pi\over2}$, while the $D6'_-$-brane is obtained
by rotating the $\bar D6'$-brane by an angle ${\alpha\over2}=-{\pi\over2}$.
The conjugation of $S_{45}$ in passing from the $D6'$ to the $\bar D6'$-brane
is due to the different GSO projection on branes and anti-branes.}

\noindent
Some comments are in order at this point:
\item{(1)} The above analysis is valid as long as $j=|m|-1>-{1\over2}$
satisfies the unitarity bound \limj, \ie\ for $\alpha>\alpha_c={2\pi\over k}$.
For $\alpha<\alpha_c$, the lowest lying normalizable states belong to
the continuum ($j=-{1\over2}+is$) and the physics is different.

\item{(2)} An interesting question is whether $m$, \mappar, is quantized
on the cigar. Before separating the $NS5$-branes on a circle, it is clear
that $\alpha$ in \vfoursixpr\ (and hence $m$) is arbitrary -- the angle
between the $D4$ and $D6'$-branes is not constrained. After the separation
the situation is different. The $D$-branes are wrapped around the cigar, and
the momentum around the cigar $m$ appears to be quantized, $m\in Z$. For
$D$-branes wrapped around a circle one can shift the momentum by an arbitrary
fractional amount, by turning on a Wilson line of the gauge field on the
branes (around the circle). However, on the cigar there are no
non-contractible cycles, and hence one expects the quantization of
$m$ to persist.

\lref\afkoun{I. Antoniadis, S. Ferrara and C. Kounnas, hep-th/9402073,
Nucl. Phys. {\bf B421} (1994) 343.}%
\lref\egsu{T. Eguchi and Y. Sugawara, hep-th/0002100.}

\item{(3)} The quantization of $m$ can be
seen in other ways as well. One is to use a T-dual description,
in which the cigar is replaced by $N=2$ Liouville \refs{\gk,\afkoun,\egsu},
and the $D$-branes are at points on the dual $S^1$. The $N=2$ Liouville
superpotential effectively pins down the (T-dual) field $Y$ \jtbos\
to points on the circle when one goes far down the $N=2$ Liouville throat
to the strong coupling region. Since the $D$-branes extend into the
throat and lie at points (independent of $\phi$) on the circle, their
positions on the circle must coincide with those determined by the
superpotential. Another way to see the quantization is to note that
quantization of $\alpha$ has a natural geometric interpretation when the
branes are ending on fivebranes which lie on a circle. Since the $D4$
and $D6'$-branes point towards the center of the circle on which the
fivebranes lie, the angle between the $D$-branes, $\alpha\over2$, must
be an integer multiple of $2\pi\over k$ -- the angular separation between
the $k$ fivebranes. This leads to the same conclusion, \mappar, $m\in Z$.
Thus, the $D$-branes seem to know that the smooth cigar is in fact associated
with $k$ fivebranes on a circle.

\item{(4)} It is interesting to compare the symmetry properties of
\agrec\ to those expected of $Q$, $\tilde Q$.
The vertex operator  $V_{4-6'}^+$ is
charged under the unbroken rotation symmetry $SO(2)_{45}\times
SO(2)_{89}$ of the configuration of fig. 10. The $SO(2)_{45}$
charge is carried by the spin field $S_{45}$ in \agrec.
In units in which the supercharges have $SO(2)_{45}$ charge
$\pm{1\over2}$, $V_{4-6'}^+$ has charge ${1\over2}$.
The $SO(2)_{89}$ charge of \vfoursixpr\ is carried by the
last factor, $\exp(-i{\alpha\over2\pi}\sqrt{k\over2}Y)$.
Normalizing it such that the supercharges again have charge
$\pm{1\over2}$, one finds that the charge of $V_{4-6'}^+$ is
$-{k\alpha\over4\pi}$. Therefore, the $SO(2)_{89}$ charge
of \agrec\ is $-k/4$.
{}From the discussion above one learns that $\tilde Q$
carries the same charge as $Q$.
These assignments are similar but not identical to those postulated
in brane theory in the past. In our notation, the $SO(2)_{45}\times
SO(2)_{89}$ charge of $Q, \tilde Q$ was postulated to be $({1\over2},0)$
(see \eg\ \gkrev, discussion between eqs. (182) and (183)).
In the present construction, we find charge
$({1\over2},-{k\alpha\over4\pi})$ for $Q$ and $\tilde Q$.
The discrepancy in the $SO(2)_{89}$ charges might be related
to that found in \gk; a better understanding of its origin is
left for future work.

\noindent
For $\alpha<\pi$ the lowest lying state \alphmass\ is tachyonic;
the stable vacuum is obtained by its condensation. This process
has a very natural interpretation from the point of view of brane
theory, which also makes it clear what is the endpoint of the
condensation.
For $\alpha\not=\pm \pi$, the configuration of fig. 10 is not
supersymmetric, hence stability needs to be checked. For $\alpha<\pi$
the configuration of fig. 10 can reduce its energy by having the
$D4$-brane slide away from the $NS5$-brane so that it ends on
the $D6'$-brane instead (fig. 11).
The resulting vacuum is stable. For $\alpha>\pi$, the configuration
of figs. 10, 11 is stable under small deformations. Indeed, the lowest lying
open string state is massive in this case \alphmass.

\vskip 1cm
\centerline{\epsfxsize=45mm\epsfbox{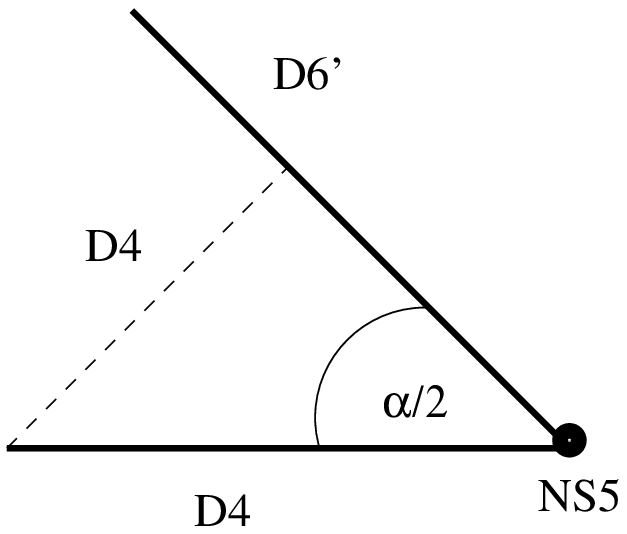}}
\centerline{Fig.\ 11}
\vskip .5cm

{}From the point of view of the brane configurations describing four
dimensional gauge theories like that of fig. 3, one can change
$\alpha$ from $\pi$ in a number of ways.  One is to change the
relative position of the $NS5$ and $NS5'$-branes in $x^7$ by the
amount $\Delta x^7$. In
the gauge theory on the $D4$-branes this corresponds to turning on
a Fayet-Iliopoulos D-term \refs{\hanawit, \gkrev}. Depending on
the sign of the FI term, either $Q$ or $\tilde Q$ should condense
to minimize the D-term potential
\eqn\dtermp{V_D\sim (Q^\dagger Q-\tilde Q^\dagger\tilde Q-r)^2}
where $r$ is proportional to $\Delta x^7$. The gauge theory analysis
is nicely reproduced by our string theory considerations\foot{Both in
gauge theory and in string theory, in the context of the full configuration
of fig. 3 the foregoing discussion is valid for $N_f\ge N_c$ (see
\gkrev\ for details).}. $\Delta x^7>0$
corresponds to $0<\alpha<\pi$. In this case the ground state of the
$4-6_+$ string is tachyonic, $Q$ condenses and the $D4$-branes
detach from the $NS5'$-branes and attach to the $D6$-branes.
$\Delta x^7<0$ corresponds to $\alpha>\pi$ for the $4-6_+$
string which is hence massive, but since $0<-\alpha<\pi$ for
the $4-6_-$ strings ($\tilde Q$), a similar process of condensation
to the above occurs for them.

Another way of changing $\alpha$ is to tilt the $D6$-branes
by some angle in the $(x^6, x^7)$ plane, which must lead to a
potential similar to \dtermp.

\noindent
More comments on the supersymmetric case $\alpha=\pi$:

\item{(1)} Like in \gk, for values of $j$ that satisfy the unitarity
constraint \limj, the full spectrum obtained from \disrepp, \mmmaaa\
and its generalization to other observables is non-tachyonic.
\item{(2)} The excited twist field \exctwist\
does not create from the vacuum massless states;
this is shown in appendix B.
\item{(3)} Massive states
form hypermultiplets, as they should; this is also shown in appendix B.
\item{(4)} Like in the closed string case \refs{\abks,\gk}, there is a
continuum of $\delta$-function normalizable $4-6'$ string states
corresponding to $j\in -1/2+i\IR$. These states are separated by an
energy gap of order $1/l_s$ from the massless discrete states
discussed above.

\subsec{$D4$-branes ending on $NS5$-branes from opposite sides}

We next turn to the configuration of fig. 4b. A stack of $N_L$
$D4$-branes ends from the left on $k$ coincident $NS5$-branes,
and a stack of $N_R$ $D4$-branes ends on the $NS5$-branes from
the right. In this case, in the parametrization \sutwo,
the $D4$-branes on the left intersect the group manifold at $g=1$
while the ones on the right intersect it at $g=-1$ (see fig. 12).

\vskip 1cm
\centerline{\epsfxsize=140mm\epsfbox{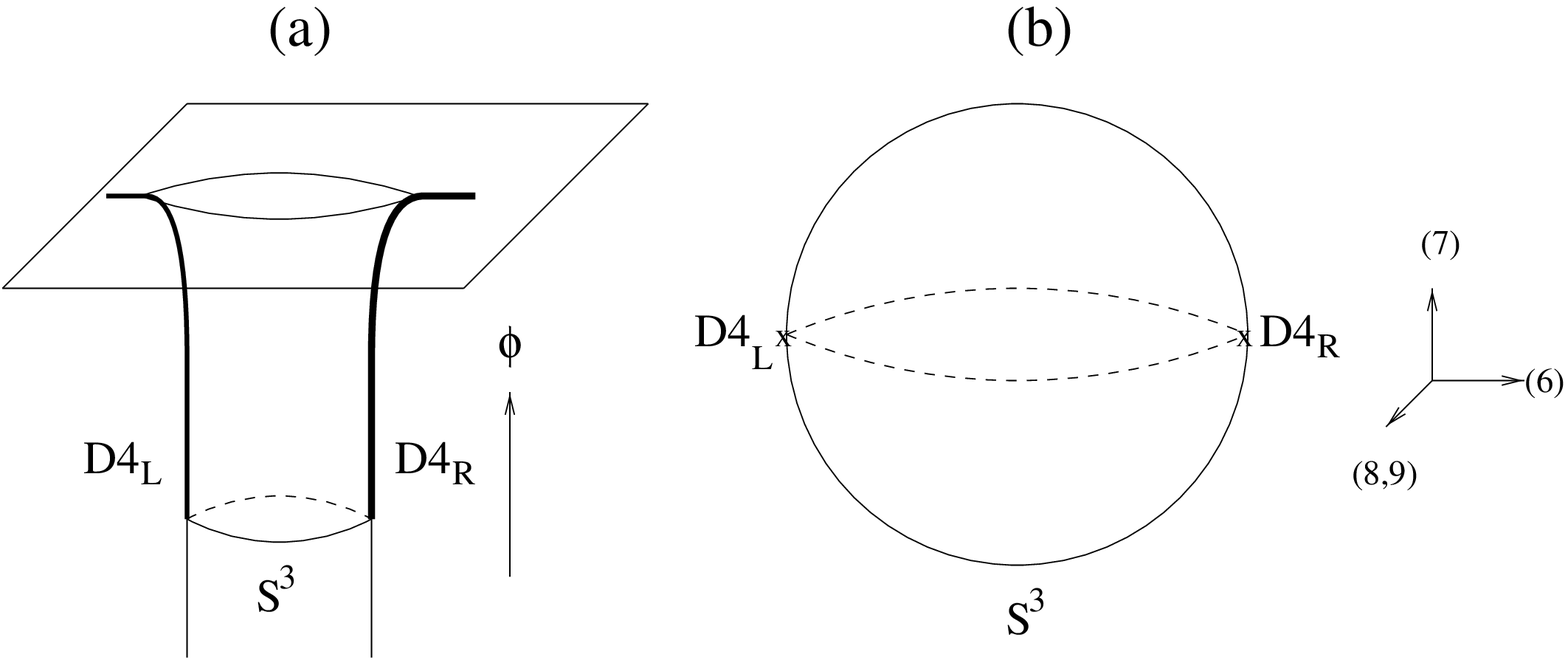}}
\centerline{Fig.\ 12}
\vskip .5cm

We first consider the case where the two stacks of $D4$-branes are
at the same point in the $(x^4,x^5)$ plane.
To construct the vertex operator for emitting the lowest lying
$4_L-4_R$ string in the geometry of fig. 12, we can follow the
discussion of the previous subsection. The dependence on
$(x^0,x^1,x^2,x^3,\phi)$ is the same as in the $4-6'$ case.
In the $(x^4,x^5)$ directions we now have Dirichlet-Dirichlet
boundary conditions. Therefore, the $4-4$ vertex operator is:
\eqn\agvff{V_{4-4}=e^{-\varphi}e^{ik_\mu x^\mu}e^{\beta\phi}V_2}
The contribution $V_2$ of the $SU(2)$ SCFT to \agvff\ is the same
as in section {\it 5.1}, with $\alpha=\pm 2\pi$:
\eqn\agvtwo{V_2^{\pm}=\sigma_{\mp 2\pi}=e^{\pm i \sqrt{k\over 2}Y}}
Hence
\eqn\agvffpm{V_{4-4}^{\pm}=e^{-\varphi}e^{ik_\mu x^\mu}e^{\beta\phi}
e^{\pm i \sqrt{k\over 2}Y}}
To describe $V_{4-4}^{\pm}$ in the cigar background \backgr, we use
eqs. \cylcig, \mappar, with $\alpha=\mp 2\pi$. The vertex operators
\agvffpm\ become in this geometry:
\eqn\agbtoj{V_{4-4}^{\pm}=e^{-\varphi}e^{ik_\mu x^\mu}
V_{j,\pm {k\over 2}}}
The mass shell condition reads:
\eqn\agmsc{{k_\mu^2\over 2}+{k\over 4}-{j(j+1)\over k}=\frac12
\Rightarrow
-k_\mu^2={2\over k}\Big[({k\over 2}-1){k\over 2}-j(j+1)\Big]}
As in \gk\ and section {\it 5.1}, the two point function of
\agbtoj\ has a series of poles corresponding to discrete
representations of $SL(2)$, and the lowest lying state
corresponds to
\eqn\aglls{j=|m|-1={k\over 2}-1}
Plugging \aglls\ into \agmsc\ we find that in this case
$V_{4-4}^{\pm}$ create massless states from the vacuum.

The vertex operators which couple to these massless particles
take the form \agbtoj, \aglls:
\eqn\agffrec{V_{4-4}^\pm(k_\mu^2=0)=e^{-\varphi}e^{ik_\mu x^\mu}
V_{{k\over 2}-1,\pm {k\over 2}}}
As expected from gauge theory, $V_{4-4}^{\pm}$ describe two complex
scalar fields $Q$, $\tilde Q^\dagger$, in the $(N_L,N_R)$
of the $U(N_L)\times U(N_R)$
symmetry on the $D$-branes, respectively. The complex conjugates
$Q^\dagger$, $\tilde Q$ arise from $4_L-4_R$ strings with the
opposite orientation.

In this case, the charges of $Q,\tilde Q$ under the $SO(2)_{45}
\times SO(2)_{89}$ R-symmetry are $(0, {k\over2})$. The $SO(2)_{45}$
agrees with expectations. The $SO(2)_{89}$ charge predicted by
brane theory is $1/2$, which is the value obtained by extrapolating
our results to $k=1$ (which, as discussed above, is in fact outside
the range of validity of our approach). For $k>1$, the situation is
less clear, but it is worth pointing out that in addition to the fact
that the same issue already arose in the previous subsection and in
\gk, it also appeared in brane theory before. As reviewed in \gkrev,
the theory on $D4$-branes ending on $k$ $NS5$-branes typically
contains chiral superfields $\Phi$ with a polynomial superpotential
$s_0\Phi^{k+1}$. The coupling $s_0$ appears to be charged under the
analog of $SO(2)_{89}$. The resolution of all these problems is
again left for future work.

\noindent
Comments:

\item{(1)}
If the $N_L$ $D4$-branes are located at $x^4=x^5=0$, and
the $N_R$ $D4$-branes are at $(x^4,x^5)=(a,b)$, one can
apply the discussion of section {\it 4.1}.
As in \flverop, the vertex operators $V_{4-4}^{\pm}$ have to
be multiplied by the ``winding'' generating factor
$\exp\{{i\over \pi}[a(x_L^4-x_R^4)+b(x_L^5-x_R^5)]\}$.
The lowest lying $4_L-4_R$ states have mass squared
${1\over \pi^2}(a^2+b^2)$, in agreement with expectations.

\item{(2)}
For the case $\alpha=\pm 2\pi$, the $SU(2)$ current
algebra is untwisted (see \alphacurbcthree).
The discussion of section {\it 4.2} shows that the
$D4_L$ and $D4_R$ branes correspond in this case to
the boundary states with $j=0$ and $j={k_B\over2}$,
respectively. Hence, the strings that connect the
two transform in the $j={k_B\over2}$ representation
of the bosonic $\widehat{SU(2)}$. Denoting the corresponding
vertex operators by $\sigma_{{k_B\over2},m}$
($m=-{k_B\over2},-{k_B\over2}+1,\cdots, {k_B\over2}-1, {k_B\over2}$),
we find that $V_2$ in \agvff, \agvtwo\ is given by
\eqn\agtform{V_2^\pm=e^{\pm i\sqrt{k\over 2}Y}
=\chi^{\pm}\sigma_{\mp 2\pi}^B=
\chi^{\pm}\sigma_{{k_B\over2},\pm {k_B\over2}}}
The last expression can be thought of as the highest and
lowest weight states in a spin ${k\over2}={k_B\over2}+1$
representation of the total $SU(2)$. The full set of primaries
at the lowest level of $4_L-4_R$ strings is the following:
\eqn\fullset{\eqalign{
(i)\;&e^{-\varphi}e^{ik_\mu x^\mu}e^{\beta\phi}\left(
\chi\cdot\sigma_{k_B\over2}\right)_{j={k\over2}}\cr
(ii)\;&e^{-\varphi}e^{ik_\mu x^\mu}e^{\beta\phi}\left(
\chi\cdot\sigma_{k_B\over2}\right)_{j={k\over2}-2}\cr
(iii)\;&e^{-\varphi}e^{ik_\mu x^\mu}e^{\beta\phi}\left[
\sum_Ma_M\psi^M\sigma_{k_B\over2}+a_8\left(
\chi\cdot\sigma_{k_B\over2}\right)_{j={k\over2}-1}
\right]\cr
}}
In $(i)$ and $(ii)$, $\chi$ and $\sigma_{k_B\over2}$
are coupled to a representation of total spin ${k\over2}$
and ${k\over2}-2$, respectively. $(iii)$ contains six physical
combinations (after imposing the physical state conditions
on the polarization coefficients $a_M$, and eliminating null
states); $\{\psi^M\}=\{\psi^\mu, \psi^4, \psi^5, \chi^r\}$.
It can be shown that out of the operators \fullset\ only
those whose $SU(2)$ part is
\agtform\ create massless particles (see appendix C).

\item{(3)}
One can study $4_L-4_R$ strings when the angle between the two
kinds of branes is generic, $\alpha/2$. For $\alpha/2\not=\pm\pi$,
an analysis similar to the one in section {\it 5.1} shows that
the lowest lying $4_L-4_R$ states are tachyonic. For $\alpha=0$,
one finds a system of parallel $D4-\bar D4$.
This is discussed in the next subsection.

\subsec{Rotating $D4-D4$ systems into $D4-\bar D4$}

The vertex operators creating the low lying states of $4_L-4_R$
strings when the angle between the $D4_L$ and $D4_R$-branes is
$-\pi<\alpha/2<\pi$
are (see subsection {\it 5.1}):
\eqn\agda{V_{4-4}(\alpha/2)=
e^{-\varphi}e^{ik_\mu x^\mu}e^{\beta\phi}
e^{- i {\alpha\over 2\pi} \sqrt{k\over 2} Y}}
Repeating the analysis of sections {\it 5.1}, {\it 5.2},
one finds that the lowest lying states that \agda\ creates
from the vacuum are tachyonic.
At $\alpha=0$, when the two $D4$-branes are anti-parallel,
\eqn\agdb{V_{4-4}(0)=e^{-\varphi}e^{ik_\mu x^\mu}e^{\beta\phi}}
is the tachyon vertex operator creating the low lying state of a string
connecting a $D4$-brane parallel to an anti-$D4$-brane ($\bar D4$-brane),
both ending on the $NS5$-branes from the same side; it is a $4_L-\bar 4_L$
string.

\lref\sen{A. Sen, hep-th/9904207.}

As before, the tachyon signals that the $D4$-branes could reduce their
energy by reconnecting. Imagine for simplicity that the ends of the
$D4$-branes are pinned down far from the $NS5$-branes (see fig. 13).
Tachyon condensation corresponds to a process where the $D4$-branes
(solid lines) connect to each other and detach from the fivebrane.
They can then reduce their energy by stretching straight between
the two pinned endpoints (dashed line). As is clear from figure 13,
their energy decreases in the process. As $\alpha\to 0$, the fourbranes
annihilate (or more generally become lower dimensional branes). This
generalizes the results of \sen\ on brane annihilation to $D$-branes
ending on fivebranes.

\vskip 1cm
\centerline{\epsfxsize=60mm\epsfbox{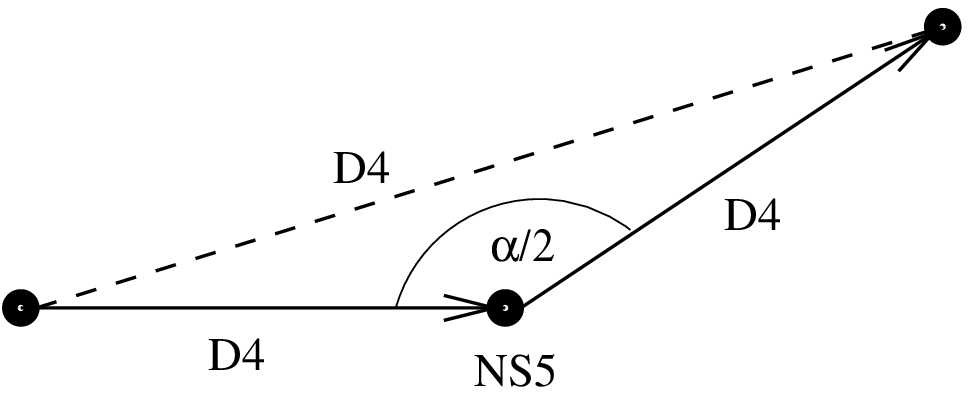}}
\centerline{Fig.\ 13}
\vskip .5cm

Similarly, the vertex operator creating the low lying states
of $4_L-4_L$ strings when the angle between the $D4_L$-branes is
$\alpha/2$ is
\eqn\agdc{V_{4-4} (\alpha/2)=
e^{-\varphi}e^{ik_\mu x^\mu}e^{\beta\phi}
e^{i [({\alpha\over |\alpha|}-{\alpha\over 2\pi})H
- \sqrt{k_B\over 2}{\alpha\over 2\pi}u]}}
(the last exponent is the excited twist field \exctwist).
The extra factor of $\exp(\pm iH)$ in \agdc\ compared to
\agda\ is due to the different signs of the GSO projection
in the two cases resulting from the reversed orientation of
the $D4_R$-brane. At $\alpha=0$, we obtain the $4_L-4_L$ vertex
operator which in the cigar geometry is (recall \boschipm\ and
see appendix B for notation)
\eqn\agdd{V^\pm_{4-4}(0)=e^{-\varphi}e^{ik_\mu x^\mu}e^{\beta\phi}\chi^\pm
\rightarrow
e^{-\varphi}e^{ik_\mu x^\mu}z_1^{k-2}V_{j,\pm 1}}
(the $\pm$ in \agdd\ is correlated with $\alpha\rightarrow 0_+$ or
$\alpha\rightarrow 0_-$ in \agdc).
At first sight one might be puzzled why for two parallel
$D4$-branes ending on a stack of $NS5$-branes one does not find
additional massless particles. For example, in section 2 it
was argued that fundamental strings stretching between the
fourbranes give rise to massless non-Abelian gauge bosons
on the fourbranes. The reason we are not supposed to see those
here is that we are studying the physics of the $D$-branes
in the near-horizon geometry of the fivebranes. Only states
whose wavefunctions are bound to the fivebranes can give rise
to normalizable modes in our analysis. The wavefunctions
of the gauge bosons discussed in section 2 are in contrast
spread out in $x^6$ and, in particular, are not localized at
the fivebranes. From the point of view of the near-horizon
geometry, they are non-normalizable.

At $\alpha/2=\pm\pi$ \agdc\ turns into
\eqn\agde{V_{4-4}(\pm\pi)=e^{-\varphi}e^{ik_\mu x^\mu}e^{\beta\phi}
\sigma^B_{\pm 2\pi}}
which is the tachyon vertex operator for a $4_L-\bar 4_R$ string
connecting a $D4$-brane and a $\bar D4$-brane located on opposite
sides of the stack of $NS5$-branes (see fig. 14).

\lref\mstong{S. Mukhi, N.V. Suryanarayana and D. Tong, hep-th/0001066,
JHEP {\bf 0003} (2000) 015.}%
\lref\mukhis{S. Mukhi and N.V. Suryanarayana, hep-th/0003219.}%

\vskip 1cm
\centerline{\epsfxsize=70mm\epsfbox{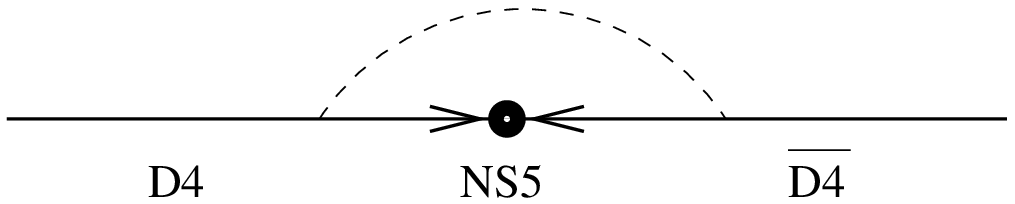}}
\centerline{Fig.\ 14}
\vskip .5cm

Brane constructions involving $D4$ and $\bar D4$-branes
ending on $NS5$-branes were recently studied in \refs{\mstong,
\mukhis}. The present work can be used to shed more light on
such configurations.

\subsec{A $D6$-brane intersecting a stack of $NS5$-branes}

We now turn to the configuration of fig. 4c.
A stack of $D4$-branes ends from the left on $k$
coincident $NS5$-branes, and a $D6$-brane intersects
the $NS5$-branes at the same value of $(x^4,x^5)$
as the fourbranes.

{}From the brane geometry point of view in Type IIA,
the notion of putting a $D6$-brane on top of $k$
{\it coincident} $NS5$-branes is not well defined,
due to the HW transition \hanawit.
For instance, for a single $NS5$-brane one must specify if
the $D6$-brane is located to the left of the $NS5$-brane or
to its right. In the first case, the $D6$-brane can be
moved freely away from the $NS5$-brane to the left in the
$x^6$ direction, but if moved away to the right an extra
$D4$-brane is created, stretched between the $D6$-brane
and the $NS5$-brane. In the second case, the $D4$-brane
will be created when moving the $D6$-brane to the left.
The configuration space of this system is thus separated
into two disconnected components.

Similarly, for the case of $k$ $NS5$-branes one has to specify
which of these fivebranes are to the left of the $D6$-brane and
which of them are located to its right. If $n\leq k$  fivebranes
are to the left of the $D6$-brane, $n$ $D4$-branes will be created
when the $D6$-brane moves to the left away from the stack of
$NS5$-branes, and $k-n$ $D4$-branes will be formed if it is taken
away to the right. Thus the configuration space consists of $k+1$
sectors -- there are $k+1$ possibilities of what is meant by placing
a $D6$-brane on top of $k$ coincident fivebranes.

\lref\yoshi{Y. Yoshida, hep-th/9711177,
Prog. Theor. Phys. {\bf 99} (1998) 305.}%

This ambiguity in type IIA string theory can be understood in M-theory
on $\IR^{10}\times S^1$ when the radius of the $S^1$ is large in
Planck units \yoshi. The $D6$-brane becomes a Kaluza-Klein monopole,
\ie\ a bundle whose fiber is the eleventh circular coordinate
$x^{10}$ of radius $R_{10}$.
The bundle is non-trivial and two patches are needed to trivialize it.
These patches can be chosen as the two
halves of the ten dimensional space, one with $x^6\geq 0$
to the right of the sixbrane and the other with $x^6\leq 0$
to its left. Denote by $x^{10}_{\pm}$ the fiber coordinate
over the two different patches. If $\vartheta$
is the azimuthal angle in the $(x^4,x^5)$ plane,
{\it i.e.} $\tan(\vartheta )= {x^5\over  x^4}$,
then the transition between the fiber coordinates in the two
patches on the transition $(x^4,x^5)$ plane at $x^6=0$ is:
${x^{10}_+\over R}={x^{10}_-\over R_{10}} + \vartheta$.
{}From eleven dimensional standpoint, an $NS5$-brane is an
$M5$-brane which is a point in $x^{10}$, stretching along
the $(x^4,x^5)$ plane.
Starting with $k$ $NS5$-branes in presence of a $D6$-brane,
one must specify which of these $M5$-branes have
a definite $x^{10}_+$ coordinate, and which have a well
defined $x^{10}_-$ coordinate.

The division of configuration space into sectors is visible also
in the worldsheet CFT description of section 4. Unlike the
$D6'$-brane case studied in section {\it 5.1}, the $D6$-brane intersects
the group manifold \sutwo\ not at a point but along a two-sphere of
constant $x^6$. This corresponds to the conjugacy class $C_\theta$
\symbc, where $\theta$ satisfies
$\cos({\theta \over 2}) ={x^6\over |\vec x|}$.
In section 4 we have seen that consistent boundary conditions of this
type exist only for $k-1$ values of $\theta$ given by eq. \qtheta.
The $D6$-brane is thus allowed to intersect the stack of $k$ $NS5$-branes
only at fixed quantized values of $x^6$, namely, on two-spheres with
quantized sizes $S^2_n$, $n=1,2, \dots,k-1$ (see fig. 15).

\vskip 1cm
\centerline{\epsfxsize=60mm\epsfbox{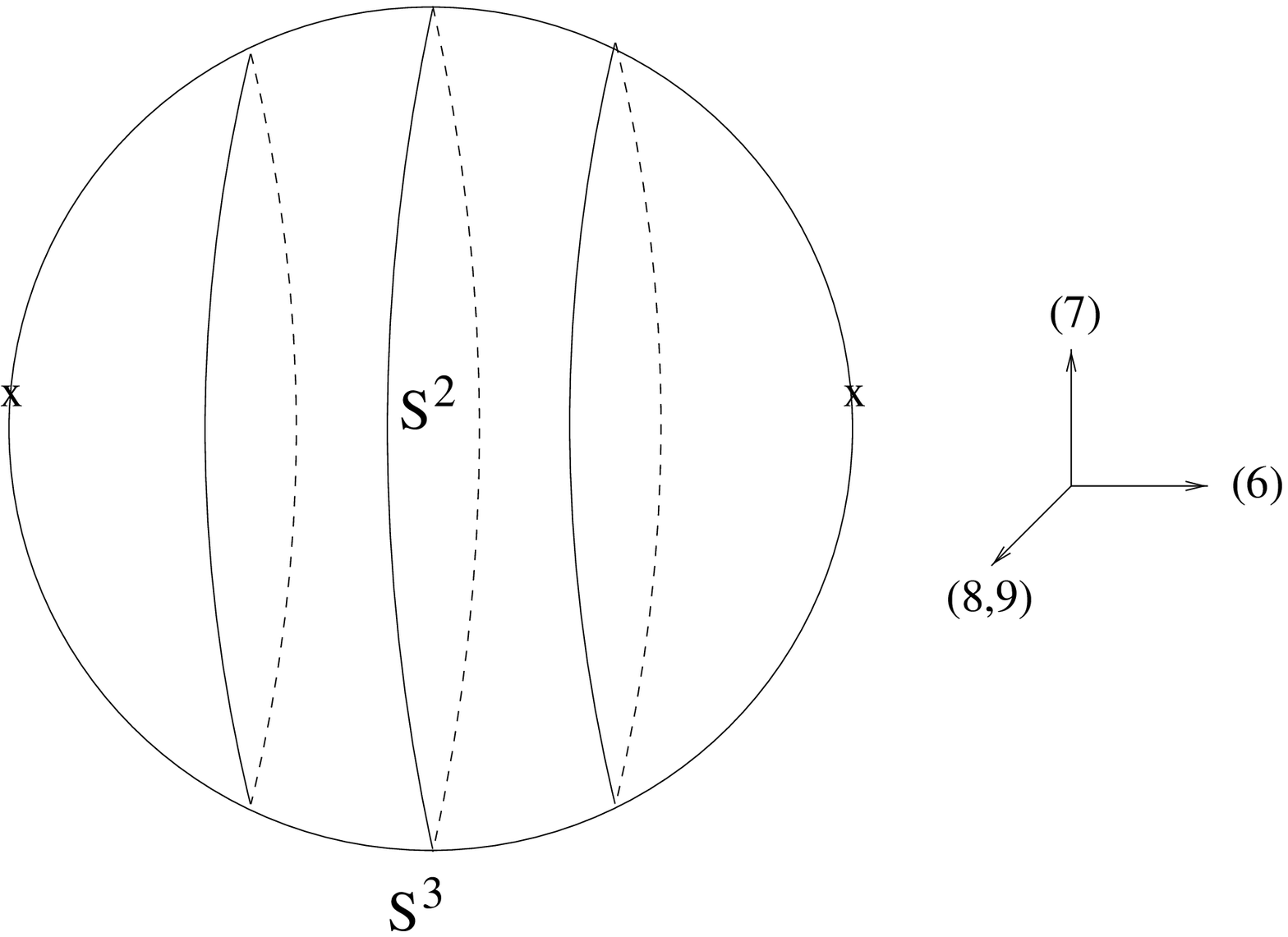}}
\centerline{Fig.\ 15}
\vskip .5cm

It is natural to identify these $k-1$
different boundary states with the various possible positions of
the $D6$-brane among the $k$ $NS5$-branes, discussed above. There
we had $k+1$ possible states, two of which are absent in the CFT
approach. A possible interpretation of this fact is that when the
$D6$-brane is either to the right or to the left of all $NS5$-branes,
it is not bound to the fivebranes and hence does not give rise to
a boundary state in the CHS geometry.

Anyhow, we can again study the massless $4-6$ open strings when the
$D6$-brane is characterized by $\theta = 2\pi {l\over k_B}$ in \qtheta,
with $2l=0,1,\cdots,k_B$. Such a string connects the $g=1$ boundary
state with $l=0$, corresponding to the $D4$-brane, to a spin $l$
boundary state corresponding to the $D6$-brane (see fig. 16). The
$4-6$ string belongs to representations contained in the fusion of
spin $0$ and spin $l$, \ie\ the spin $l$ representation. This consists
of $2l+1$ operators $\sigma_{lm}$, $m=-l,-l+1,\cdots,l-1,l$,
with scaling dimension
\eqn\diml{h(\sigma_{lm})={l(l+1)\over k}}
As in subsection {\it 5.2},
we get for each $l$ the vertex operators:
\eqn\agfk{V_{lm}=e^{-\varphi}e^{ik_\mu x^\mu}e^{\beta\phi}
\left(\chi\cdot \sigma_l\right)_{l+1,m}}
and some additional operators that do not create massless states.
The two operators $V_{lm}$ \agfk\ with $m=\pm (l+1)$ do give
rise to poles at $k_\mu^2=0$ in amplitudes and thus create massless
particles of charge $\pm(l+1)$.

\vskip 1cm
\centerline{\epsfxsize=60mm\epsfbox{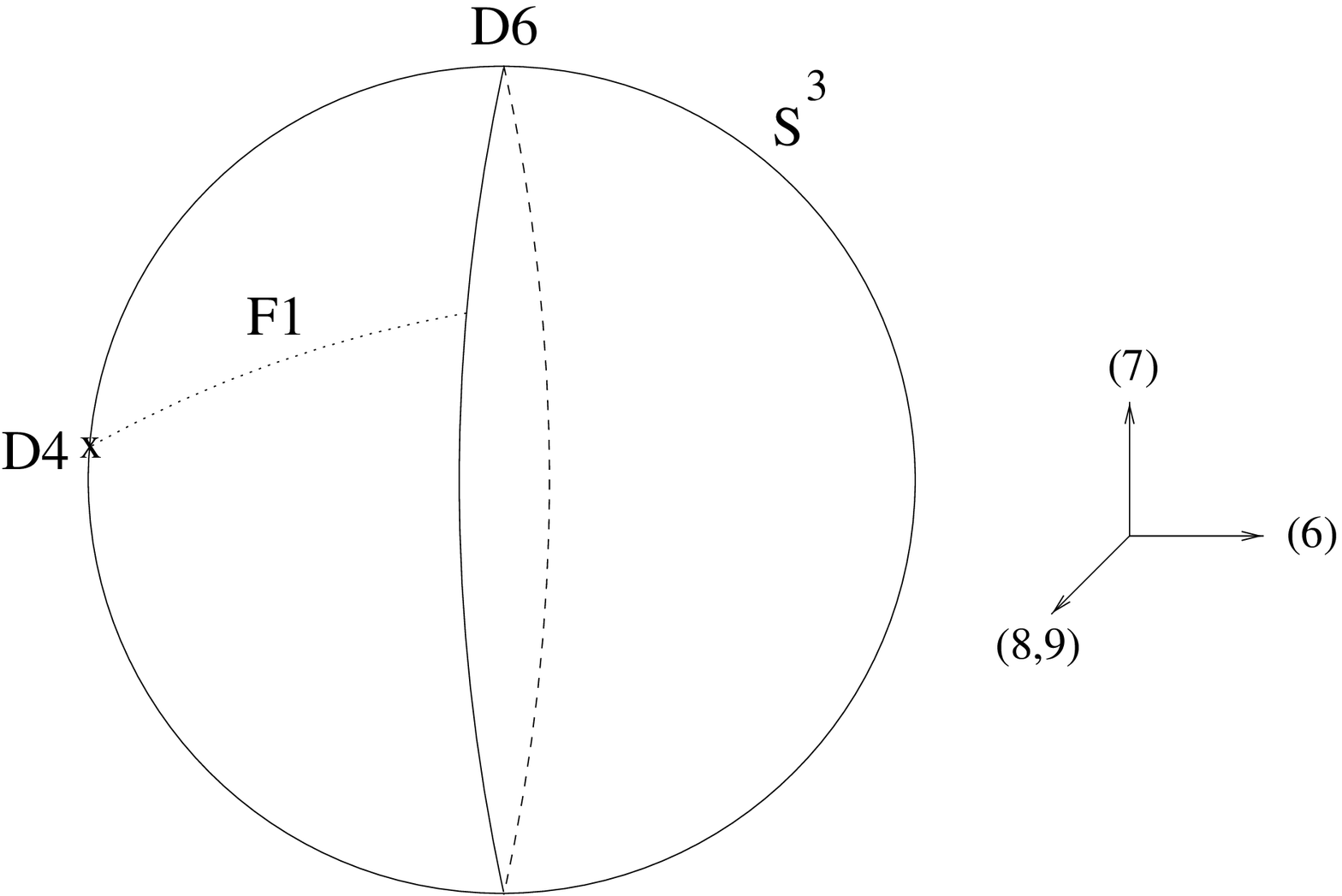}}
\centerline{Fig.\ 16}
\vskip .5cm

Recently, it was argued that $n$ $D0$-branes in an $SU(2)$
WZW background, which are $n$ points on $S^3$, can turn
into a single $D2$-brane which wraps a two-sphere with a quantized
radius $S^2_n$ \refs{\bdsnew,\arsnew}.
This meshes nicely with the discussion above.
In the near horizon of $k$ $NS5$-branes, a $D4$-brane is a point on
$S^3$ while a $D6$-brane wraps an $S^2_n\in S^3$, $n=1,2,\dots, k-1$.
Hence, $n$ $D4$-branes can condense into a single $D6$-brane
in the $\theta=\pi {n-1\over k-2}$ conjugacy class \qtheta, namely,
a $D6$-brane wrapping the sphere $S^2_n$. This is the HW transition.
The $D6$-brane on $S^2_n$ is located, say, to the left of $n$ out of
the $k$ $NS5$-branes. Moving it to the right, past these $n$ fivebranes,
creates $n$ $D4$-branes stretched from the $D6$-brane to each of the $n$
fivebranes. In the near horizon limit these look like $n$ points on $S^3$.

\bigskip
\noindent{\bf Acknowledgements:}
We thank M. Berkooz, A. Hanany, N. Itzhaki, Y. Oz and O. Pelc
for useful discussions. We also thank the ITP in Santa
Barbara for hospitality during the course of this work.
S.E. thanks the IAS in Princeton. This research was
supported in part by NSF grant \#PHY94-07194 and by the
Israel Academy of Sciences and Humanities -- Centers of
Excellence Program. The work of A.G. and E.R. is also
supported in part by BSF -- American-Israel Bi-National
Science Foundation. D.K. is supported in part by DOE grant
\#DE-FG02-90ER40560.

\appendix{A}{$\langle V_{jm} V_{j,-m}\rangle$ on the disc}

We first present some useful formulae:
\eqn\agaz{C(a,b)=\int_{-\infty}^{\infty} |x|^{a-1}|1-x|^{b-1} dx
=B(a,1-a-b)+B(a,b)+B(1-a-b,b)}
Here $B$ is the Euler beta function:
\eqn\agazz{B(a,b)=\int_0^1 x^{a-1}(1-x)^{b-1} dx=
{\Gamma(a)\Gamma(b)\over \Gamma(a+b)}}
where
\eqn\agazzz{\Gamma(a+1)=a\Gamma(a)~, \qquad
\Gamma(-n+\epsilon)={(-)^n\over n!\epsilon}+O(1)~, \quad n=0,1,2,\dots}
To find the analytic structure of the two point functions
of boundary operators discussed in this work, like \vvff,
we need to compute the two point functions
$\langle V_{jm} V_{j,-m}\rangle$ on the disc.
This is the purpose of this appendix.

The calculation is similar to its closed string analog
\refs{\teschner,\gk}, which we follow below.
The CFT on the cigar is obtained by the coset construction
from that on $AdS_3$. The natural observables in CFT on the
Euclidean version of $AdS_3$ on the disc are functions
$\Phi_j(x;z)$ which transform as primaries under
the diagonal (left $+$ right) $SL(2)$ current algebra
(see section 4.2). $x$ is an auxiliary real variable.

The two point function of $\Phi_j$ is\foot{Here
and below we suppress the dependence of correlation functions
on the worldsheet location $z$
on the real line, the boundary of the upper half plane.}
\eqn\agaa{\langle\Phi_j(x)\Phi_j(x')\rangle
=A(j,k)|x-x'|^{-2(j+1)}}
where $A(j,k)$ is an analytic function of $j$ in the domain
\limj; its precise form will not play a role below.

To study the coset it is convenient to ``Fourier transform''
the fields $\Phi_j(x)$ and define the mode operators
\eqn\agab{\Phi_{jm}=\int_{-\infty}^{\infty} \Phi_j(x) |x|^{j+m} dx}
The two point functions of the modes $\Phi_{jm}$ are equal to those
of the $SL(2)/U(1)$ coset theory:
\eqn\agac{\langle V_{jm} V_{j,-m}\rangle=
\langle \Phi_{jm} \Phi_{j,-m}\rangle}
Using \agac, \agab\ and \agaa\ one finds that
\eqn\agad{\langle V_{jm} V_{j,-m}\rangle=A'(j,k)
\int_{-\infty}^{\infty} |y|^{j-m}|1-y|^{-2(j+1)} dy=C(j-m+1,-2j-1)}
Using eqs. \agaz, \agazz, \agazzz\ one can finally express the
two point functions in terms of gamma functions as:
\eqn\agae{\eqalign{&\langle V_{jm} V_{j,-m}\rangle=A'(j,k)\times\cr
&\left(
{\Gamma(j-m+1)\Gamma(j+m+1)\over \Gamma(2j+2)}+
{\Gamma(j-m+1)\Gamma(-2j-1)\over \Gamma(-j-m)}+
{\Gamma(j+m+1)\Gamma(-2j-1)\over \Gamma(-j+m)}
\right)}}
Using eq. \agazzz\ we can now find the analytic structure of
$\langle V_{jm} V_{j,-m}\rangle$: it has single poles for $j,m$
satisfying \disrepp.\foot{For $j=m$ one finds that
the three terms in \agae\ conspire to give $0$.
This implies that the extra poles at $j=m$, found in
some special cases in \gkunpub,
appear on the sphere but not on the disc.}

\appendix{B}{}

In this appendix we show that the excited twist field \exctwist\ does
not give rise to massless particles even for $\alpha=\pi$, and in the
case $\alpha=\pi$ we verify that the massive states created by
\lowsig\ and \exctwist\ have degeneracy two, and are hence
organized into hypermultiplets.

The twist field \exctwist\ has a conformal weight
\eqn\agbb{h(\sigma'_\alpha)=
h(\sigma_\alpha)+\frac12\Big(1-{\alpha\over\pi}\Big)
={k\over 4}\Big({\alpha\over 2\pi}\Big)^2
+\frac12\Big(1-{\alpha\over\pi}\Big)}
and a $J^3_{\rm total}$ charge (recall \jtbos, \yuh)
\eqn\agbc{m(\sigma'_\alpha)=m(\sigma_\alpha)+1
=-{k\over 2}{\alpha\over 2\pi}+1}
Its decomposition on $SU(2)/U(1)\times U(1)$ thus reads
\eqn\agbd{\sigma'_\alpha=z_1^{k-2}
e^{-i\sqrt{2\over k}({k\over 2}{\alpha\over 2\pi}-1)Y}}
where $z_1^{k-2}$ (the notation will become clear soon) is an
operator in the $N=2$ minimal model $SU(2)_k/U(1)$ with
\eqn\agbe{h(z_1^{k-2})=h(\sigma'_\alpha)
-{2\over k}{({k\over 2}{\alpha\over 2\pi}-1)^2\over 2}
={1\over 2}{k-2\over k}}
It can be shown that $z_1^{k-2}$ is a chiral operator in the $N=2$
minimal model with a $U(1)_R$ charge $R(z_1^{k-2})=(k-2)/k$;
it is the highest charge operator in the chiral ring of $SU(2)_k/U(1)$,
$\{z_1^i|i=0,1,\dots,k-2\}$, $R(z_1^i)=i/k$.

Collecting the above, recalling \vfoursixpr\ (with $\sigma_\alpha$
replaced by $\sigma'_\alpha$) and \cylcig, \mappar, \vvff, we find that
the $4-6'$ vertex operator under consideration, and what it becomes
in the cigar geometry \backgr, is
\eqn\agbf{V'_{4-6'}=e^{-\varphi}\sigma_{45}S_{45}e^{ik_\mu x^\mu}
e^{\beta\phi}\sigma'_\alpha \rightarrow
e^{-\varphi}\sigma_{45}S_{45}e^{ik_\mu x^\mu}
z_1^{k-2}V_{j,-{k\over 2}{\alpha\over 2\pi}+1}}
The lowest lying state corresponds to $n=1$ in \disrepp, namely,
$j=|m|-1$; its mass is
\eqn\agbg{(M'_\alpha)^2=\frac12+{2\over k}j
=\frac12-{2\over k}+\Big|{\alpha\over 2\pi}-{2\over k}\Big|}
Since $j>-1/2$ \limj, we have
\eqn\agbg{(M'_\alpha)^2>\frac12-{1\over k}\geq 0}
for any $\alpha$ allowed in the unitarity range.

A particular case is the supersymmetric $\alpha=\pi$ configuration.
In this case, the twist operators \lowsig\ and \exctwist\ take the form
\eqn\agbh{\eqalign{\sigma_\pi^-&=e^{-{i\over 2}H}
e^{-{i\over 2}\sqrt{k_B\over 2}u}=e^{-{i\over 2}\sqrt{k\over 2}Y}
\cr
\sigma_\pi^+&=e^{{i\over 2}H}e^{-{i\over 2}\sqrt{k_B\over 2}u}
=z_1^{k-2}e^{-i\sqrt{2\over k}({k\over 4}-1)Y}}}
The corresponding $4-6'$ vertex operators turn in the cigar geometry
\backgr\ into
\eqn\agbi{\eqalign{
V_{4-6'}^-&=e^{-\varphi}\sigma_{45}S_{45}e^{ik_\mu x^\mu}
e^{\beta\phi}\sigma_\pi^- \rightarrow
e^{-\varphi}\sigma_{45}S_{45}e^{ik_\mu x^\mu}V_{j,m_-}\cr
V_{4-6'}^+&=e^{-\varphi}\sigma_{45}S_{45}e^{ik_\mu x^\mu}
e^{\beta\phi}\sigma_\pi^+ \rightarrow
e^{-\varphi}\sigma_{45}S_{45}e^{ik_\mu x^\mu}z_1^{k-2}V_{j,m_+}
}}
where
\eqn\agbj{m_\pm=-{k-2\over 4}\pm \frac12}
Physical states in spacetime are obtained when $j$ in $V_{4-6'}^\pm$
is \disrepp\
\eqn\agbk{j^n_\pm = |m_\pm|-n~, \qquad n=1,2,\dots}
respectively. Notice that\foot{Below we restrict to the case $k\geq 4$;
for $k=2,3$ there are no massive excitations in the range \limj.}
\eqn\agbl{j_+^n=j_-^{n+1}}
The on-shell conditions $h(V_{4-6'}^\pm)=1$ imply
\eqn\agbm{{(M_\pm^n)^2\over 2}={k-4\over 16}-{j^n_\pm(j^n_\pm+1)\over k}}
Using \agbj, \agbk, \agbl, \agbm, one finds
\eqn\agbn{(M_-^n)^2={1\over k}(n-1)(k-2n)}
\eqn\agbo{(M_+^n)^2=(M_-^{n+1})^2={1\over k}n(k-2-2n)}
The unitarity bound \limj\ restricts $n$ to a certain range.
In this range $(M_+^n)^2>0$ for any $n$, while $(M_-^n)^2\geq 0$
for any $n$ and equality is satisfied iff $n=1$.
We thus see \agbo\ that there is a degeneracy two for
all massive states; they get organized into
$4d$ hypermultiplets $(Q,\tilde{Q})$, as they should.
The only massless state $M_-^1=0$ \agbn\ is non-degenerate; it corresponds
to the chiral superfield $Q$ considered in section {\it 5.1}.

\appendix{C}{}

In this appendix we show that, except for \agtform,
the other operators in \fullset\ do not create massless states
from the vacuum.

The operators in \fullset\ are linear combinations of
\eqn\agfkind{V_m^a=e^{-\varphi}e^{ik_\mu x^\mu}e^{\beta\phi}
\chi^a\sigma_{{k_B\over 2},m}~, \qquad a=\pm,3}
and
\eqn\agskind{V_m^M=e^{-\varphi}e^{ik_\mu x^\mu}e^{\beta\phi}
\psi^M\sigma_{{k_B\over 2},m}~, \qquad M=\mu,4,5,r}
The operators $\sigma_{{k_B\over 2},m}$ in
\agfkind\ and \agskind\ carry $m$ units of charge under $J^3$
and have scaling dimension
\eqn\agca{h(\sigma_{{k_B\over 2},m})={1\over k}{k_B\over 2}({k_B\over 2}+1)=
{k-2\over 4}}
The decomposition of $\sigma_{jm}$ on $SU(2)/U(1)\times U(1)$
reads
\eqn\agcb{\sigma_{jm}=V'_{jm}e^{i\sqrt{2\over k}mY}}
where $V'_{jm}$ is an operator in the $SU(2)/U(1)$
SCFT with scaling dimension
\eqn\agcc{h(V'_{jm})={j(j+1)-m^2\over k}}
Now, going from the cylinder $(\phi,Y)$ to the $SL(2)/U(1)$ cigar
we take
\eqn\agcd{e^{\beta\phi}e^{i\sqrt{2\over k}mY}\rightarrow V_{jm}}
(see \cylcig, \mappar).
Altogether, in the background \backgr, the operators \agskind\ take the form
\eqn\agcd{V^M_m=e^{-\varphi}e^{ik_\mu x^\mu}e^{\beta\phi}
\psi^M\sigma_{{k_B\over 2},m}\rightarrow
e^{-\varphi}e^{ik_\mu x^\mu}\psi^M V'_{{k_B\over 2},m}V_{jm}}
When $k_\mu^2=0$, the on-shell condition for $V^M_m$ reads
\eqn\agce{{1\over k}{k_B\over 2}({k_B\over 2}+1)-{j(j+1)\over k}=0
\Rightarrow j={k_B\over 2}}
Since $|m|\leq {k_B\over 2}$, the condition \disrepp\ is not satisfied.
Hence, no massless particle is emitted by $V^M_m$.
Similarly, $V_m^3$ in \agfkind\ does not create
massless states from the vacuum.

The operators $V_m^\pm$ in \agfkind\ include a factor
$\chi^\pm\sigma_{{k_B\over 2},m}$ which decomposes as
\eqn\agcf{\chi^\pm\sigma_{{k_B\over 2},m}=
\Psi_m e^{i\sqrt{2\over k}(m\pm 1)Y}}
where $\Psi_m$ is in the $SU(2)/U(1)$ SCFT and has
\eqn\agcg{h(\Psi_m)={k\over 4}-{(m\pm 1)^2\over k}}
In the geometry \backgr, the operators $V_m^\pm$ now take the form
\eqn\agch{V_m^\pm=e^{-\varphi}e^{ik_\mu x^\mu}e^{\beta\phi}
\chi^\pm\sigma_{{k_B\over 2},m}\rightarrow
e^{-\varphi}e^{ik_\mu x^\mu}\Psi_m V_{j,m\pm 1}}
When $k_\mu^2=0$, the on-shell condition gives again
$j={k_B\over 2}$ \agce.
Since $m\pm 1$ take values between $-{k_B\over 2}\pm 1$
and ${k_B\over 2}\pm 1$,
only the operators $V^+_{k_B\over 2}$ and $V^-_{-{k_B\over 2}}$
-- the operators in \fullset\ whose $SU(2)$ part is \agtform\ --
satisfy \disrepp.
Therefore, the other operators $V_m^{\pm}$ in \agfkind\ do not
emit massless particles.

\listrefs
\end